\documentclass[aps,showpacs,twocolumn,superscriptaddress,prb]{revtex4}
\usepackage{graphicx,color,epsfig,rotate}
\usepackage{amssymb,amsmath,bm}

\begin{document}

\title{Thermal Conductivity in Large-$J$ Two-Dimensional Antiferromagnets:
Role of Phonon Scattering}
\author{A. L. Chernyshev} 
\affiliation{Department of Physics and Astronomy, University of California, Irvine, California 92697, USA}
\author{Wolfram Brenig} 
\affiliation{Institute for Theoretical Physics, Technical University Braunschweig, D-38106 Braunschweig, Germany}
\date{\today}
\begin{abstract} 

Motivated by the recent heat transport experiments in 2D
antiferromagnets, such as La$_2$CuO$_4$, where the exchange coupling $J$ is
larger than the Debye energy $\Theta_{\rm D}$, we discuss different types
of relaxation processes for magnon heat current with a particular
focus on coupling to 3D phonons. We study thermal conductivity by these
in-plane magnetic excitations using two distinct techniques, 
Boltzmann formalism within the relaxation-time approximation and 
memory-function approach. Within these approaches, a close
consideration is given to the scattering of magnons by both acoustic and
optical branches of phonons. A remarkable accord between the two methods
with regards to the asymptotic behavior of the effective relaxation rates
is demonstrated. Additional scattering mechanisms, due
to grain boundaries, impurities, and finite correlation length in the
paramagnetic phase, are discussed and included in the calculations of the
thermal conductivity $\kappa(T)$. Again, we demonstrate a close
similarity of the  results  from the two techniques of calculating
$\kappa(T)$. Our
complementary approach strongly suggests that scattering from optical or
zone-boundary phonons is important for magnon heat current
relaxation in a high temperature window of $\Theta_D\lesssim T \ll J$.

\end{abstract}
\maketitle

\section{Introduction}
\label{intro}

After almost three decades of intensive studies, cuprates continue to
attract significant interest because of their outstanding properties and
due to  the  continued research effort in high-temperature
superconductivity.\cite{Ronnow15,Kastner98,ScalapinoRMP,Davis,Vojta,Basov,Armitage}  
In particular, their magnetic properties remain
inspirational, both as potentially responsible for the mechanism of
superconductivity\cite{ScalapinoRMP} and on their own right as relevant to
a larger class of low-dimensional antiferromagnets and as a test case for
various theoretical models. \cite{Ronnow15,Greven} However, understanding
of some of the properties of the magnetic excitations in layered
cuprates remains incomplete. This concerns interactions of such
excitations with themselves and various other perturbations, as well as the
role of such interactions in observable spectroscopic and transport
phenomena.

Earlier studies of thermal conductivity in La$_2$CuO$_4$
\cite{Hess03,others} have demonstrated large contribution of
magnetic excitations to the in-plane thermal transport, thus offering a
unique window into their properties which are not easily accessible by
other methods.  More recent experimental advances \cite{Hess15} call for
a deeper theoretical insight into  the mechanisms of magnon heat
current dissipation. This interest goes beyond a particular material and 
highlights  a  broader importance of   general understanding of the transport
phenomena in a wider class  of antiferromagnets.
While scattering of magnons among themselves and due to fluctuations of the
order-parameter in the paramagnetic state has received significant
attention in the past,\cite{Chubukov} the  impact  of phonons  on
magnon  lifetime and other properties of magnetic excitations has
only recently begun receiving attention.\cite{Ronnow14}

In this context, the current work focuses on the physics
of magnon scattering and its role in  magnetic thermal transport of
quantum antiferromagnets. While our approaches are generic, our analysis is 
strongly motivated by 
La$_2$CuO$_4$ and related cuprates. Therefore, we concentrate on
the case of 2D magnons and superexchange coupling $J$ large compared
to the Debye energy $\Theta_D$, for temperatures $T\lesssim 0.4J$, where
the validity of a magnon description is well-established \cite{Takahashi}. 
Moreover, most of our work is devoted
to the magnon-phonon scattering to clarify the role
of this less-studied relaxation mechanism. Additional effects, such as
grain boundary and impurity scattering, and the role of finite correlation length
are discussed less extensively.

In addition to investigation of the transport properties of layered cuprates, we
also contribute to the formal development of transport theory by
contrasting the results from two complementary methods, the Boltzmann
theory and the memory function technique.  For the relaxation-time
approximation within the Boltzmann approach, we take advantage of the large
energy scale of magnetic excitations and thus operate with asymptotic,
long-wavelength expressions augmented by appropriate cut-offs and some
well-justified modeling of the optical phonon spectra.  For the
magnon-phonon scattering we advocate the use of a simplified ``effective
phonon DoS'' approach, which allows for straightforward yet fairly
realistic calculations.  In the memory-function approach, on the other
hand, we maintain microscopic expressions for the magnon spectra valid in
the entire Brillouin zone, while using coupling to a single dispersive
phonon branch with a model dispersion, which introduces scattering on both
acoustic and optical-like zone-boundary modes.  The  final  results
 from  both approaches, i.e.  the  thermal conductivities
vs. temperature, are  found to be  remarkably similar.   Moreover, a
complete agreement between both
approaches on the power-law asymptotic regimes, controlled by the optical and acoustic phonons, is also
demonstrated.

 The paper is organized as follows:  we begin with a general discussion
of the qualitative features of the magnon-phonon scattering in
Sec.~\ref{qualitative} followed by intuitively clear details of Boltzmann
approach and calculations of $\kappa(T)$ within it in Sec.~\ref{boltzmann}.
We continue with the exposition of the memory-function approach and its
results for the effective relaxation rates and thermal conductivity in
Sec.~\ref{MF}.  A brief discussion of other scattering mechanisms and of
their respective roles is given in the corresponding sections devoted to
the thermal conductivity calculations. Technical details, discussion of the
physical range of spin-phonon coupling, etc., are delegated to several
Appendices.

\section{Model and qualitative considerations}
\label{qualitative}

Generally speaking, a spin system on a lattice can always be described by a
Hamiltonian  consisting  of spin-only and lattice-only parts, ${\cal
H}_{\rm s}$ and ${\cal H}_{\rm ph}$  respectively ,  in addition to
a coupling between them, which we will assume to be  of magnetoelastic
nature ${\cal H}_{\rm s-ph}$
\begin{equation}
{\cal H}={\cal H}_{\rm s}+{\cal H}_{\rm ph}+{\cal H}_{\rm s-ph}={\cal H}_{0}+{\cal H}_{\rm s-ph}\, .
\label{0}
\end{equation}
Having in mind La$_2$CuO$_4$ and related cuprates, we take a simple,
nearest-neighbor-dominated Heisenberg model with the superexchange constant
$J$ on a square lattice to be a faithful description of the 2D
antiferromagnet of interest. Appendix \ref{app0} offers the standard linear
spin-wave treatment of it leading to the free-magnon Hamiltonian
\begin{equation}
{\cal H}_{\rm s}\Rightarrow\sum_{\bf k}\varepsilon_{\bf k}\beta^\dag_{\bf k}\beta^{\phantom\dag}_{\bf k}\,, 
\label{0s}
\end{equation}
where $\varepsilon_{\bf k}$ is the magnon energy and $\hbar\!=\!k_B\!=\!1$ from now on.

The full phonon spectrum of La$_2$CuO$_4$ is representative of that of the
other cuprates and  comprises  three acoustic and eighteen optical
modes. With  a  typical bandwidth of each branch from 50K to 400K, together
they span the range of energies reaching 900K. \cite{Pintschovius91} While
in what follows we will model them in a simplified fashion, one can write
their Hamiltonian as
\begin{equation}
{\cal H}_{\rm ph}\Rightarrow\sum_{{\bf q},\ell}\omega_{{\bf q},\ell}a^\dag_{{\bf q}\ell}a^{\phantom\dag}_{{\bf q}\ell}\,, 
\label{0ph}
\end{equation}
where $\ell$ numerates branches of phonon excitations.

A straightforward derivation yields the lowest-order magnon-phonon coupling in the following general 
form (see Appendix~\ref{appA} for details)  
\begin{eqnarray}
&&{\cal H}_{\rm s-ph}=\sum_{{\bf k, k', q}} \sum_{\ell}
\Big\{V^{\ell}_{\bf k, k', q} \beta_{\bf k'}^\dag \beta_{\bf k} 
\label{Hsph2_0}\\
&& \phantom{{\cal H}_{\rm s-ph}=}
+\frac12\, V^{{\rm od},\ell}_{\bf k, k', q}\left( \beta_{\bf k'}^\dag \beta_{-\bf k}^\dag+\mbox{H.c.} \right)\Big\}
\left(a^\dag_{{\bf q}\ell}+a_{-{\bf q}\ell}\right), 
\nonumber
\end{eqnarray}
where $V^{\ell}_{\bf k, k', q}$ and 
$V^{{\rm od},\ell}_{\bf k, k', q}$ are 
the ``normal'' and ``anomalous'' spin-phonon coupling vertices.  
For the coupling to acoustic and optical (or zone-boundary) 
phonons, they assume different asymptotic forms, discussed in the next Section and in Appendix~\ref{appA}.

Since we are interested in the thermal transport by magnons, the following
generic consideration is useful.  Because the phonon Debye energy
($\Theta_D\approx 400$K) is much smaller than the magnon bandwidth ($2.2J
>3000$K), phonons can be assumed to be in thermal equilibrium, i.e.,
playing the role of a ``bath''.  This is well-justified for temperatures
comparable to or above  half of  the phonon Debye energy
$T\agt\Theta_D/2$, which is about 200K in most cuprates.  Then neither the
momentum nor the energy of a magnon is conserved in the processes of
magnon-phonon scattering, or, in other words, the momentum and energy are
transferred from the magnon flow to the phonon bath.  In that case, magnon
relaxation time and transport times can be treated as the same.

Thus, in contrast to the inter-magnon scattering,   the magnon-phonon
scattering channel is free from many restrictions of the former 
and does not have the limitations on the phase space inherent to an Umklapp
scenario. While the spin-lattice coupling constant may be small, 
phonons at temperatures $T>200$K are abundant in the materials of interest.

Another qualitative consideration, elaborated on in Sec.~\ref{boltzmann},  is that 
magnetic excitations are confined to lower dimensions than  phonons (i.e., 2D vs 3D).  
In this case, momentum of phonons perpendicular
to the 2D planes is not conserved, which also leads to fewer restrictions on
the kinematics of the magnon-phonon scattering.  

\section{Boltzmann approach}
\label{boltzmann}

\subsection{Relaxation rates}
\label{tau}
 
\begin{figure}[t]
\includegraphics[width=0.9\columnwidth]{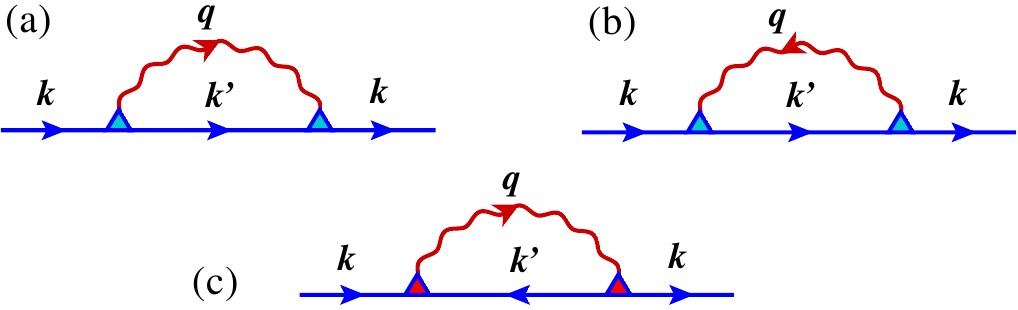}
\caption{Magnon-phonon scattering diagrams. Solid lines are magnons, wavy lines are phonons.
(a) Decay (phonon emission), (b) recombination (phonon absorption), (c) anomalous process 
(absorption of magnon). } 
\label{diagrams}
\vskip -0.3cm
\end{figure} 

 Within the Boltzmann approach, the key element is the calculation of the scattering rates. As is
 argued above,  transport and quasiparticle 
 relaxation rates of magnons due to scattering on phonons should be the same. 
Then, in the lowest order of spin-phonon coupling, the problem is reduced
to evaluation of the   diagrams in Fig.~\ref{diagrams}, 
which yield magnon relaxation rate by the standard diagrammatic  method
\begin{eqnarray}
&&\frac{1}{\tau_{\bf k}}=\pi \sum_{\bf k'}\sum_{{\bf q}_\parallel}\sum_{q_\perp}
\label{1tau}\\
&&\phantom{\frac{1}{\tau_{\bf k}}}
\Big\{\left|V_{{\bf k},{\bf k'},{\bf q}}\right|^2\left(n_{\bf q}+n_{\bf k'}+1\right)\delta_{{\bf k},{\bf k'}+{\bf q}_\parallel}
\delta\left(\varepsilon_{\bf k}-\varepsilon_{\bf k'}-\omega_{\bf q}\right)
\nonumber\\
&&\phantom{\frac{1}{\tau_{\bf k}}}
+\left|V_{{\bf k'},{\bf k},{\bf q}}\right|^2\left(n_{\bf q}-n_{\bf k'}\right)
\delta_{{\bf k'},{\bf k}+{\bf q}_\parallel}
\delta\left(\varepsilon_{\bf k'}-\varepsilon_{\bf k}-\omega_{\bf q}\right)
\nonumber\\
&&\phantom{\frac{1}{\tau_{\bf k}}}
+\left|V^{\rm od}_{{\bf k},{\bf k'},{\bf q}}\right|^2\left(n_{\bf k'}-n_{\bf q}\right)
\delta_{{\bf k'}+{\bf k},{\bf q}_\parallel}
\delta\left(\varepsilon_{\bf k'}+\varepsilon_{\bf k}-\omega_{\bf q}\right)\Big\},\nonumber
\end{eqnarray}
with the first term corresponding to the diagram in Fig.~\ref{diagrams}(a), in which
phonon is emitted,  and the second to Fig.~\ref{diagrams}(b), 
in which phonon is absorbed. The third term is due to anomalous process, Fig.~\ref{diagrams}(c), in which two magnons are 
absorbed and the phonon is emitted. We drop summation over the phonon branch index $\ell$ here, thus
considering one branch of phonons at a time.

The magnon-phonon vertices in (\ref{1tau}) are the same as in Eq.~(\ref{Hsph2_0}) and 
for the first two terms they are related by a permutation of the initial and  final states of the magnon.
Note that the 2D momentum conservation in Eq.~(\ref{1tau}), explicated by the 
$\delta_{{\bf 1},{\bf 2}+{\bf 3}}$'s, concerns only the in-plane momentum of the phonon ${\bf q}_\parallel$, while the
component perpendicular to the plane, ${\bf q}_\perp$, is not conserved. 
This is natural as magnons have infinite mass in the ${\bf q}_\perp$ direction. This feature is important 
for the future consideration and we separate sums over the components of phonon momenta 
${\bf q}=({\bf q}_\parallel,q_\perp)$ in Eq.~(\ref{1tau}) explicitly.

\subsection{Approximations}
\label{prelim}

There are 
two approximations for La$_2$CuO$_4$ and related cuprates 
that follow from the fact that all relevant energies, $T$ and $\Theta_D$, are
much smaller than the magnon bandwidth, $W\approx 2.2J$:
(i) magnon energies can be linearized
\begin{eqnarray}
\varepsilon_{\bf k}\approx v|{\bf k}|,
\label{Ek}
\end{eqnarray}
because for all practical purposes $T\ll J$ and hence $|{\bf k}|\sim T/J\ll 1$,
(ii) similarly, magnon-phonon vertices for the optical and acoustic phonons are 
($V^{\rm od}_{{\bf k'},{\bf k},{\bf q}}\!\approx\!V_{{\bf k},{\bf k'},{\bf q}}$)
\begin{eqnarray}
&&V_{{\bf k},{\bf k'},{\bf q}}\approx g_{\rm sp}^{\rm opt}\sqrt{|{\bf k}||{\bf k'}|},
\label{Vopt}\\
&&V_{{\bf k},{\bf k'},{\bf q}}\approx g_{\rm sp}^{\rm ac}\sqrt{|{\bf k}||{\bf k'}|}
\cdot\frac{|{\bf q}_\parallel|}{\sqrt{|{\bf q}|}},
\label{Vac}
\end{eqnarray}
where $v$ is the magnon velocity ($v=1.158\sqrt{2}J$, lattice constant $a=1$), and 
both $g_{\rm sp}^{\rm opt}$ and $g_{\rm sp}^{\rm ac}$ are ${\cal O}(J)\!<J$
(see Sec.~\ref{Sec_ph_sum} and Appendices~\ref{appA} and \ref{appD} for details).\cite{Bramwell90} 
For the acoustic case, the phonon dispersion in the vertex 
is also linearized, in line with the Debye approximation.

While magnon-phonon vertices in (\ref{Vopt}) and (\ref{Vac}) can be proposed on general grounds, 
they can also be derived from realistic microscopic models of spin-phonon coupling, an exercise deferred to 
Appendix~\ref{appA}. The same Appendix also deals with the role of polarization of the 3D phonons in the coupling to 
spins. The asymptotic ${\bf k}, {\bf k'}, {\bf q}\rightarrow 0$ form of these microscopic vertices agrees with
(\ref{Vopt}) and (\ref{Vac}), aside from some additional angular dependence that does not affect the results. 
The typical magnitude of coupling constants $g_{\rm sp}$'s will be discussed in Sec.~\ref{Sec_ph_sum}.

We also make a note that coupling to the optical mode should typically involve a large in-plane momentum 
transfer ${\bf q}_\parallel$,
in which case a zone-boundary phonon from the nominally acoustic phonon branch is equivalent to the optical phonon.
Such processes result in the scattering of a magnon from, e.g., the branch near
${\bf k}\rightarrow 0$ to another branch near the AF ordering vector 
${\bf k'}\rightarrow {\bf Q}_{AF}$ [$=(\pi,\pi)$]. Thus, in the following we do not distinguish between the optical and the 
zone-boundary phonons.

\begin{figure}[t]
\includegraphics[width=0.5\columnwidth]{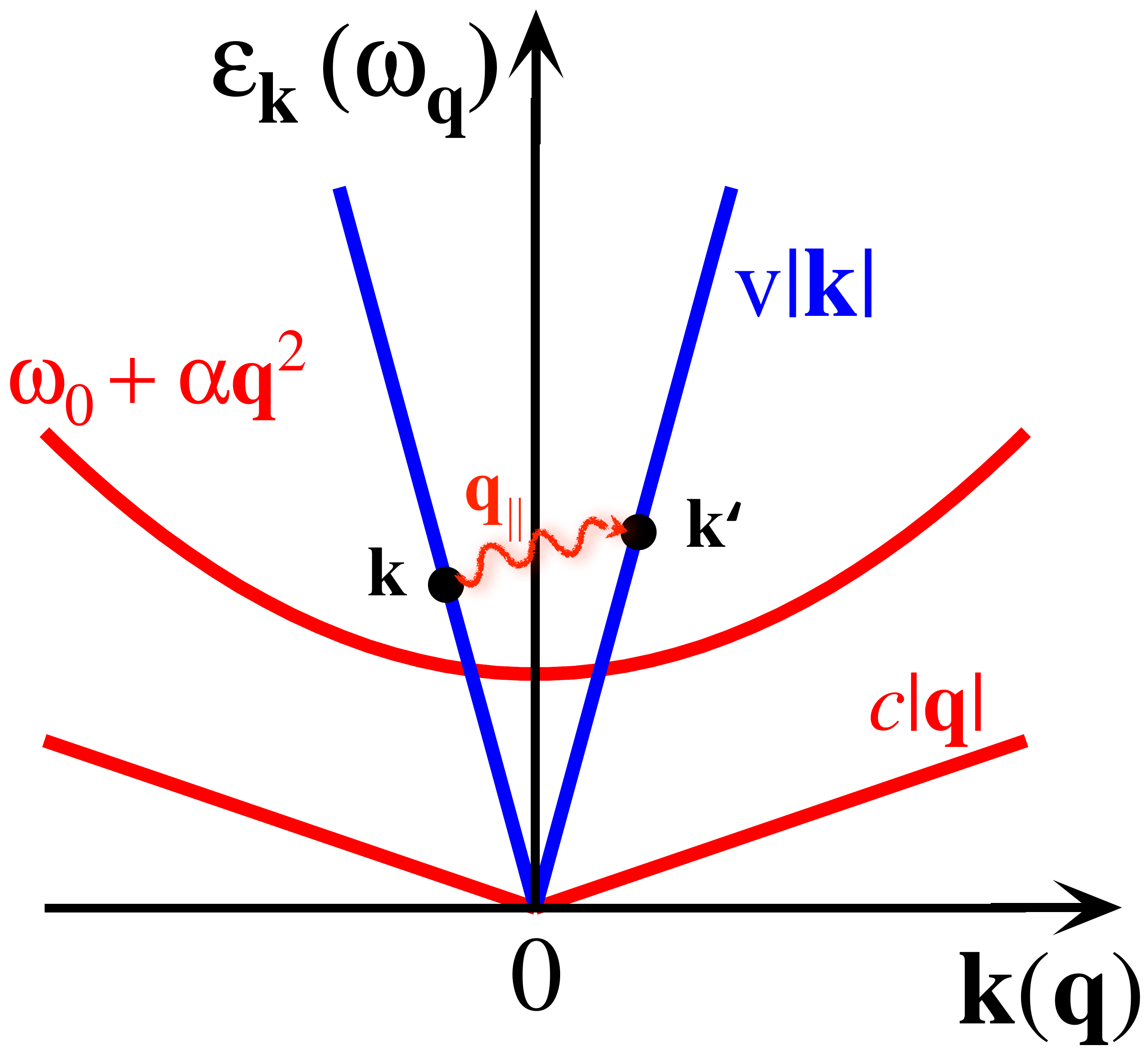}
\caption{(Color online)\ 
Qualitative sketch of the magnon ($\varepsilon_{\bf k}\!=\!v|{\bf k}|$) and acoustic and optical phonon dispersions 
($\omega_{\bf q}$'s) in the limit $v\!\gg\! c$.
Magnon and acoustic phonon energies are linearized. Schematics of the magnon [momentum 
${\bf k}$] absorbing a phonon [in-plane momentum ${\bf q}_\parallel$] is also shown.}
\label{sketch}
\vskip -0.5cm
\end{figure} 
The next line of approximations requires some qualitative kinematic consideration. 
The linearization of magnon energies in (\ref{Ek})-(\ref{Vac}) 
is well-justified for a typical $\varepsilon_{\bf k}\!\sim\!T$ and already allows for some 
simplification in  (\ref{1tau}). 
Naturally, the ``typical'' range of momenta of magnons involved in the heat transport at relevant temperatures 
is limited by $\simeq\! T/v$, which is $\ll\pi$ and concerns a small fraction of the Brillouin zone.
Then, the typical {\it in-plane component} of the phonon momenta $|{\bf q}_\parallel|\!=\!|\pm{\bf k}\mp{\bf k'}|$ 
for the phonons involved in the magnon scatterings in Fig.~\ref{diagrams} must 
also be limited to the same range, $|{\bf q}_\parallel|\!\alt\! T/v$, 
see the sketch in Fig.~\ref{sketch}. Note, that  {\it if}   (acoustic) phonons
would also be confined to 2D, this would imply that their typical energy needs to be much less than the 
energy of magnons: $\omega_{\bf q}\!\approx\! c|{\bf q}_\parallel|\!\sim\! (c/v)T\!\ll\! T$.

However, for the 3D phonons the situation is radically different. The energy conservations
in Eq.~(\ref{1tau}) imply   $\omega_{\bf q}\!=\! |v(|{\bf k}|\pm|{\bf k'}|)|$, so for the typical magnon energy $T$,
the typical energy of a phonon is also $T$. This, in turn, implies that the out-of-plane component of the phonon 
momentum is much larger than the in-plane one, $q_\perp\!\gg\!|{\bf q}_\parallel|$. 
This is particularly easy to see for the acoustic phonon, for which the combination of the (in-plane)
momentum and energy conservations yields\cite{RC}
\begin{eqnarray}
c^2\left(|{\bf k}\pm{\bf k'}|^2+q_\perp^2\right)=v^2\left(|{\bf k}|\pm|{\bf k'}|\right)^2,
\label{acoustic}
\end{eqnarray}
so for $|{\bf k}|, |{\bf k'}|, |{\bf q}_\parallel|\!\sim\! T/v$ it follows that $q_\perp\!\sim\! T/c\!\gg\! |{\bf q}_\parallel|$.
For the optical phonon, $\omega_{\bf q}\!\approx\! \omega_0\!+\! \alpha\big(\tilde{\bf q}_\parallel^2+q_\perp^2\big)$,
the argument is simply that the  typical  $|\tilde{\bf q}_\parallel|\!\sim\!T/v\!\ll\! 1$ while $q_\perp$ does 
not have such restrictions.
Note that for the optical phonon $\tilde{\bf q}_\parallel\!=\!{\bf q}_\parallel\!\pm\!{\bf Q}_{AF}$ is shifted by the 
in-plane AF-ordering vector.

Altogether, this  begs for the following approximation
\begin{eqnarray}
\omega_{\bf q}\approx\omega_{q_\perp},
\label{approx}
\end{eqnarray}
simply neglecting the dependence of the phonon energy on the (small) in-plane momentum transfer $|{\bf q}_\parallel|$.
This approximation immediately simplifies Eq.~(\ref{1tau}) as the integral over ${\bf q}_\parallel$ simply removes
the in-plane momentum delta-functions, while the rest of the expression is independent of it (see a slightly more involved
treatment of the  case of the acoustic phonon later). Then the integration over ${\bf k'}$ in (\ref{1tau}) is simply removed
by the energy conservation using linearized magnon energies.  Lastly, the remaining integration over $q_\perp$
can be rewritten by introducing an effective DoS for phonons. Thus, one of the main results 
of this work is the development of an ``effective''
approach in which ${\bf k}$- and $T$-dependencies of the relaxation rates in Eq.~(\ref{1tau}) are given by simple 1D 
integrals. A very close precision of this approach is demonstrated in Appendix \ref{appB} 
by a comparison with the direct integration in Eq.~(\ref{1tau}) without approximation of Eq.~(\ref{approx}).

\subsection{Effective phonon DoS}

 After using Eq.~(\ref{approx}), the dependence of the integrand in 
Eq.~(\ref{1tau}) on $q_\perp$ is only through $\omega_{q_\perp}$, so it is natural to introduce an 
``effective'' density-of-states of phonons
\begin{eqnarray}
D^\perp(\omega)=\sum_{q_\perp} \delta\left(\omega-\omega_{q_\perp}\right).
\label{phDos}
\end{eqnarray}
We would like to clarify that the ``effective'' DoS  is {\it not} the full phonon DoS, but a 1D
version of it, which corresponds to the DoS of phonons with the vanishing in-plane momentum. 

\subsubsection{Optical}

\begin{figure}[t]
\includegraphics[width=0.999\columnwidth]{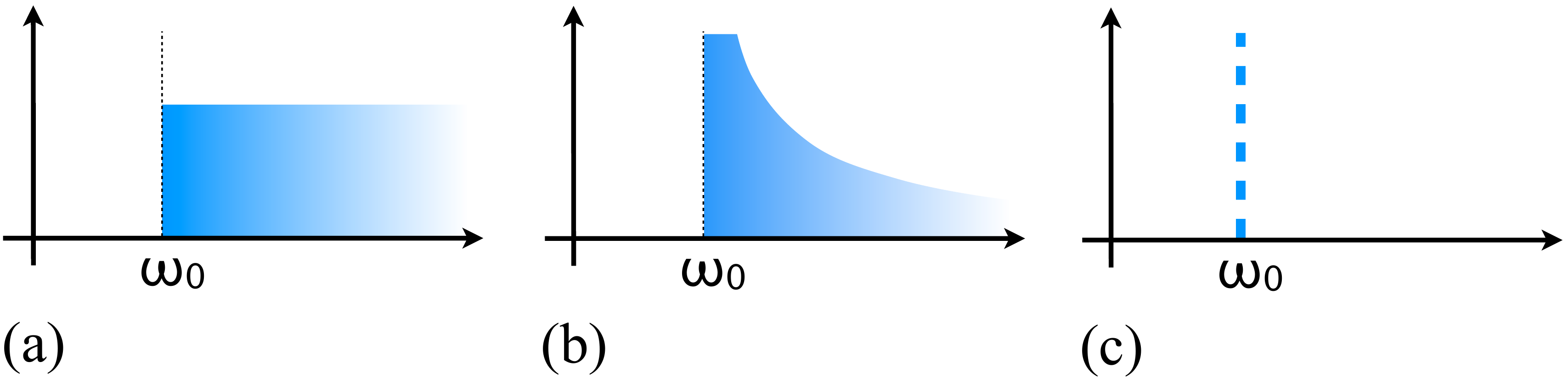}
\caption{(Color online)\ Qualitative pictures of the effective DoS for the optical phonons, 
$D^\perp(\omega)$, Eq.~(\ref{phDos}).
(a) ``Model I'' is the constant with the minimal energy $\omega_0$, (b) ``Model II'', same with a more
realistic square-root singularity at $\omega_0$, (c) ``Model III'' is 
for the ``flat'' mode  at $\omega_0$, Eqs.~(\ref{phDos1}), (\ref{phDos2}),  and (\ref{phDos3}), respectively.}
\label{phModels}
\vskip -0.4cm
\end{figure}

Given that the phonon spectrum of cuprates  
has more than a dosen of optical modes, covering the range from 
100K to 900K, \cite{Pintschovius91} we reserve the right to model them in a more straightforward fashion.
A sketch of such ``model'' densities of states is shown in Fig.~\ref{phModels}.
First model, which will be referred to as ``Model~I'', is just a constant DoS with the gap that corresponds to the lowest
optical mode
\begin{eqnarray}
D^\perp_{\rm I}(\omega)=\frac{\Theta\left(\omega-\omega_0\right)}{\omega_{\rm max}-\omega_0} .
\label{phDos1}
\end{eqnarray}
The ``Model II'', Fig.~\ref{phModels}(b), corresponds to the optical mode in the sketch in Fig.~\ref{sketch} 
with dispersion $\omega_{\bf q}\!=\!\omega_0\!+\!\alpha {\bf q}^2$ and
includes a more realistic square-root singularity at $\omega_0$ due to the 1D nature of 
the ``effective DoS'' in (\ref{phDos})
\begin{eqnarray}
D^\perp_{\rm II}(\omega)=\frac{\Theta\left(\omega-\omega_0\right)}{2\pi\sqrt{\alpha (\omega-\omega_0)}},
\label{phDos2}
\end{eqnarray}
with $\alpha\!=\!(\omega_{\rm max}\!-\!\omega_0)/\pi^2$ from normalization.
``Model III''
\begin{eqnarray}
D^\perp_{\rm II}(\omega)=\delta\left(\omega-\omega_0\right),
\label{phDos3}
\end{eqnarray}
corresponds to the ``flat'' optical mode. We would like to 
note that the results for the thermal conductivity discussed later are remarkably
insensitive to the choice of the specific model for the phonon DoS.

\subsubsection{Acoustic}
 The  effective  phonon DoS for the dispersion $\omega_{q_\perp}\!=\!c q_\perp$ is a constant, 
same as the ``Model I'' in (\ref{phDos1}), but with no gap and an upper cutoff being $\Theta_D(\sim\!c)$
\begin{eqnarray}
D^\perp_{\rm ac}(\omega)=\frac{1}{\pi c} \Theta\big(\Theta_D-\omega\big).
\label{phDos4}
\end{eqnarray}
The validity of (\ref{phDos4}) is also restricted from below by $\omega_{\rm min}=(c/v)\varepsilon_{\bf k}$ 
at which $q_\perp\!\approx\! |{\bf q}_\parallel|$, as we will discuss later.

\subsection{Scattering on the optical phonon}
\label{Sec_opt}

The case of optical phonon is straightforward since magnon-phonon coupling within the approximation 
(\ref{Vopt})  does not depend on the phonon momentum. Using linearized form of magnon dispersion and 
magnon-phonon vertex from (\ref{Ek}) and (\ref{Vopt}), approximation of Eq.~(\ref{approx}), 
and replacing $\sum_{\bf k'}$ with  $\frac{1}{\pi}\int k' dk'$  which takes into account two magnon modes per Brillouin
zone of the square lattice, yields
\begin{eqnarray}
&&\frac{1}{\tau_{\bf k}}\approx \left(\frac{g_{\rm sp}^{\rm opt}}{v}\right)^2 
\left(\frac{\varepsilon_{\bf k}}{v^2}\right)
\int_{0}^{\omega_{\rm max}} d\omega\, D^\perp(\omega)
\label{3tau}\\
&&\phantom{\frac{1}{\tau_{\bf k}}}
 \times\Big\{\Theta\big(\varepsilon_{\bf k}-\omega\big)
 \left(\varepsilon_{\bf k}-\omega\right)^2\Big(n(\omega)+n(\varepsilon_{\bf k}-\omega)+1\Big)
\nonumber\\
&& \phantom{\frac{1}{\tau_{\bf k}}\Big\{\Theta\big(\varepsilon_{\bf k}-\omega\big)}
+\left(\varepsilon_{\bf k}+\omega\right)^2\Big(n(\omega)-n(\varepsilon_{\bf k}+\omega)\Big)
\nonumber\\
&& \phantom{\frac{1}{\tau_{\bf k}}\Big\{}
+\Theta\big(\omega-\varepsilon_{\bf k}\big)
\left(\omega-\varepsilon_{\bf k}\right)^2\Big(n(\omega-\varepsilon_{\bf k})-n(\omega)\Big)\Big\},
\nonumber
\end{eqnarray}
where the first term is from the phonon-emission diagram in Fig.~\ref{diagrams}(a), the second is due to 
phonon-absorption in Fig.~\ref{diagrams}(b), and the third is the anomalous term in Fig.~\ref{diagrams}(c), 
first, second, and third terms in Eq.~(\ref{1tau}), respectively.
For the phonon-emission term the phonon energy is limited from above 
by the energy of the magnon that emits it, $\varepsilon_{\bf k}$, and in the anomalous 
term the situation is reversed as the phonon energy must exceed that of the magnon, so the integration 
is limited from below.  In (\ref{3tau}), $\omega$ is also restricted implicitly through the DoS by $\omega_0$, the 
lowest energy of the optical mode, and by $\omega_{\rm max}$, the highest energy of phonon bands. 
It is assumed that $\varepsilon_{\bf k}<\omega_{\rm max}$. Thus, after all the   
legitimate approximations discussed above are implemented, we have a compact expression of the relaxation rate in terms of 
1D frequency integrals, Eq.~(\ref{3tau}). Note that for the ``Model III'' (flat phonon mode) 
of the effective phonon DoS this integral in Eq.~(\ref{3tau})  is trivially removed and the relaxation rate 
is given by a compact analytical expression, presented in Appendix~\ref{appB}. 


In the limit of low temperatures, 
$T,\varepsilon_{\bf k} \!\ll\!\omega_0$, phonon-emission term in $1/\tau$ in (\ref{3tau}) is strictly zero as 
the magnon with $\varepsilon_{\bf k}\!<\!\omega_0$ cannot emit an optical phonon, 
and the two remaining terms are exponentially small, $\sim\!e^{-\omega_0/T}$.

For higher temperatures $T\!\agt\Theta_D$, using the hierarchy of scales 
$T\!>\!\varepsilon_{\bf k}\!>\!\omega_0$ (see  Ref.~\onlinecite{Herring}) 
and $n(\omega)\!\simeq\! T/\omega$ yields 
the asymptotics for the first two terms in  (\ref{3tau}):
\begin{eqnarray}
\frac{1}{\tau_{\bf k}^{(1)}}\propto \frac{T\varepsilon_{\bf k}^3}{v^2\omega_0}\,, \ \ \
\frac{1}{\tau_{\bf k}^{(2)}}\propto \frac{T^2\varepsilon_{\bf k}^2}{v^2\omega_0}\,,
\label{est_tau1}
\end{eqnarray}
where we used $g_{\rm sp}^{\rm opt}\sim v$.
This is valid for Model I and Model II of the effective phonon DoS discussed above, 
while for the Model III (flat phonon), $1/\tau^{(2)}$ has the same asymptotics $1/\tau^{(1)}$.
Although it is natural 
to expect that the heat is conducted largely by thermalized  magnons with a ``typical''  
$\varepsilon_{\bf k}\!\sim \!T$, this is not exactly so in our case, because the distribution of the heat conducting magnons 
``leans'' towards  lower energies as we shall discuss  later. Nevertheless, 
it is worth pointing out that ``on a thermal shell'', i.e., at $\varepsilon_{\bf k}=T$, where the memory-function 
and Boltzmann approaches can be  consistently compared with each other, both expressions in (\ref{est_tau1}) yield the same 
\begin{eqnarray}
\frac{1}{\tau^{\rm opt}}\propto  \frac{T^4}{v^2\omega_0}\,.
\label{est_tau2}
\end{eqnarray}
This coincides with the results of the memory-function approach discussed later.
We note that the phonon-emission is always subleading to the absorption  term,\cite{Herring} 
and the anomalous term is negligible at high $T$. 
\begin{figure}[t]
\includegraphics[width=0.999\columnwidth]{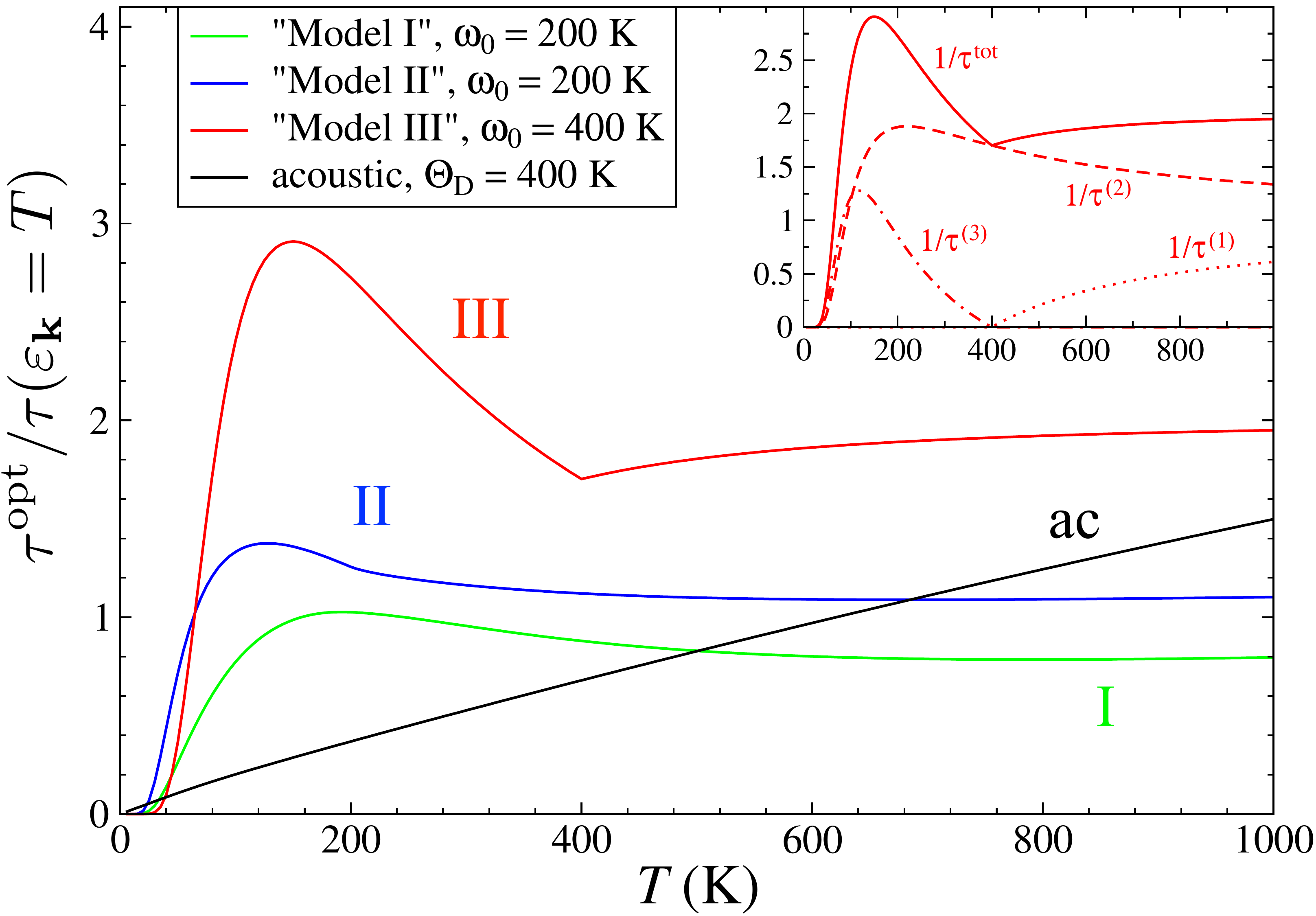}
\caption{(Color online)\ $T$-dependence of the magnon relaxation rate on the optical phonons, Eq.~(\ref{3tau}), 
for $\varepsilon_{\bf k}\!=\!T$ and using effective phonon DoS models in Eqs.~(\ref{phDos1})--(\ref{phDos3}).
The results are normalized to the high-temperature asymptotic behavior, Eq.~(\ref{est_tau2}), 
[$T^4/v^2\omega_0$]. Parameters are as discussed in text and indicated in the graph, $\omega_{\rm max}\!=\!950$K for 
Models I and II. 
The vertical axis is in units of $(g_{\rm sp}^{\rm opt}/v)^2$. Inset 
shows individual contributions of the three terms in (\ref{tauIII}) [diagrams in Fig.~\ref{diagrams}] 
for Model III. The results for the relaxation rate due to acoustic phonons, Eq.~(\ref{4tau}), are also shown 
for a representative choice of the Debye energy $\Theta_D\!=\!400$K. $J\!=\!1500$K.}
\label{Fig1tau}
\vskip -0.4cm
\end{figure} 

Our Figs.~\ref{Fig1tau} and \ref{Fig1tau_Ek} demonstrate these asymptotic trends explicitly. They 
also show a close \emph{quantitative} similarity of the magnon relaxation rates obtained from Eq.~(\ref{3tau}) 
using three different models for the effective phonon DoS, Eqs.~(\ref{phDos1})--(\ref{phDos3}).
In Fig.~\ref{Fig1tau}, the $T$-dependence is shown for the relaxation rate for
$\varepsilon_{\bf k}\!=\!T$, i.e., on the thermal shell. The results are normalized to $T^4/v^2\omega_0$ 
to make the high-temperature asymptotic behavior of (\ref{est_tau2}) apparent. The vertical axis is in 
units of $(g_{\rm sp}^{\rm opt}/v)^2$. Inset shows individual contributions of the three
terms in (\ref{3tau}) for the ``Model III'' (\ref{phDos3}). For this model, the contributions 
of the phonon-emission and anomalous terms [first and third in (\ref{tauIII})] are explicitly limited by 
the step-functions, resulting in a kink at $T\!=\!\omega_0$, while for the
other models these are smoothed out.
For the sake of a comparison with the acoustic phonons in the next Section, 
we also note that the results for the scattering on the optical phonon in Figs.~\ref{Fig1tau} and \ref{Fig1tau_Ek}
 likely underestimate the effect by a numerical factor about 2, due to 
 scattering of magnons between ${\bf k}\rightarrow 0$ and 
${\bf k}\rightarrow {\bf Q}_{AF}$ branches.

\subsection{Acoustic phonon}
\label{Sec_ac}

Below we demonstrate that the effective phonon DoS approach can be successfully extended to 
the consideration of magnon scattering off the acoustic phonons. 

The crucial difference of the acoustic phonon scattering is the form of the magnon-phonon coupling in 
Eq.~(\ref{Vac}), which leads to an extra factor $|{\bf q}_\parallel|^2/|{\bf q}|$ in the scattering
probability compared with the optical phonon case. A na\"{i}ve power-counting, together with the kinematic 
consideration in Sec.~\ref{prelim} suggest that the typical $|{\bf q}_\parallel|\!\sim\! T/v$ while 
$|{\bf q}|\!\approx\! q_\perp\!\approx\! (v/c)|{\bf q}_\parallel|$, translating this extra factor into $cT/v$ for the
relaxation rate. Using that the phonon sound velocity $c\!\sim\!\Theta_D\!\sim\!\omega_0$, this would yield the 
``thermal shell'' estimate for the relaxation rate $1/\tau^{\rm ac}\!\sim\!T^5/v^4$, to be contrasted with 
the asymptotic expression for the optical case (\ref{est_tau2}). The situation is more delicate, however, as the 
scattering in the present case is, in fact, dominated by the low-$\omega$ phonons. Technically,  the integral 
over the phonon energies diverges as $1/\omega$  and 
must be cut off at $\omega_{\rm min}\!=\!(c/v)\varepsilon_{\bf k}$, where $q_\perp$ becomes 
$\approx\!|{\bf q}_\parallel|$. Altogether, assuming $g_{\rm sp}^{\rm ac}\!\sim\!v$, 
this gives the ``thermal shell'' estimate for the acoustic case as
\begin{eqnarray}
\frac{1}{\tau^{\rm ac}}\propto  \frac{T^5}{\Theta_D v^3} ,
\label{est_tau4}
\end{eqnarray}
which is valid for both $T\!>\!\Theta_D$ and $T\!<\!\Theta_D$.
Note, that the result (\ref{est_tau4}) is the same as in the memory-function consideration, Sec.~\ref{MF}.

A  rigorous derivation of this result from Eq.~(\ref{1tau}) needs a slightly more delicate treatment 
of the magnon-phonon coupling. We use $|{\bf q}_\parallel|^2\!=\!|{\bf k'}\!-\!{\bf k}|^2$ and,
according to  approximation in Eq.~(\ref{approx}), $|{\bf q}|\!\approx\! q_\perp$.
Then the extra factor $|{\bf q}_\parallel|^2/|{\bf q}|$ in the scattering probability reads
\begin{eqnarray}
\frac{|{\bf k'}|^2+|{\bf k}|^2-2|{\bf k'}||{\bf k}|\cos\varphi}{q_\perp}\Rightarrow
\left(\frac{c}{v^2}\right)
\left(\frac{\varepsilon_{\bf k'}^2+\varepsilon_{\bf k}^2}{\omega_{q_\perp}}\right),
\label{simp_vertex}
\end{eqnarray}
where $\varphi$ is the angle between ${\bf k'}$ and ${\bf k}$ and, because of the approximation of Eq.~(\ref{approx}), 
the term with $\cos\varphi$ averages to zero upon the integration over this angle. With the result in
Eq.~(\ref{simp_vertex}) and using the effective DoS 
model for the acoustic branch introduced in Eq.~(\ref{phDos4}), we rewrite the relaxation rate in (\ref{1tau}) as
\begin{eqnarray}
\frac{1}{\tau^{\rm ac}_{\bf k}}\approx \left(\frac{g_{\rm sp}^{\rm ac}}{v}\right)^2 
\left(\frac{c\,\varepsilon_{\bf k}}{v^4}\right)
\int_{\omega_{\rm min}}^{\Theta_D} d\omega\, \frac{D_{\rm ac}^\perp(\omega)}{\omega} \, \Big\{ 
\Theta\big(\varepsilon_{\bf k}-\omega\big)\ \ \   &&
\label{4tau}\\
\times
\left(\left(\varepsilon_{\bf k}-\omega\right)^4+\varepsilon_{\bf k}^2\left(\varepsilon_{\bf k}-\omega\right)^2\right)
\Big(n(\omega)+n(\varepsilon_{\bf k}-\omega)+1\Big)&& 
\nonumber\\
+\left(\left(\varepsilon_{\bf k}+\omega\right)^4+\varepsilon_{\bf k}^2\left(\varepsilon_{\bf k}+\omega\right)^2\right)
\Big(n(\omega)-n(\varepsilon_{\bf k}+\omega)\Big)&&
\nonumber\\
+\Theta\big(\omega-\varepsilon_{\bf k}\big)&&
\nonumber\\
\times\left(\left(\omega-\varepsilon_{\bf k}\right)^4+\varepsilon_{\bf k}^2\left(\omega-\varepsilon_{\bf k}\right)^2\right)
\Big(n(\omega-\varepsilon_{\bf k})-n(\omega)\Big) \Big\},
\nonumber &&
\end{eqnarray}
where the three terms are the phonon-emission, phonon-absorption, and anomalous terms  in Fig.~\ref{diagrams}(a)-(c) 
and in Eq.~(\ref{1tau}), and $\omega_{\rm min}\!=\!(c/v)\varepsilon_{\bf k}$ as before. 

We first note that the contribution of the anomalous term in the acoustic phonon case
of Eq.~(\ref{4tau}) is by a factor of $c/v$ smaller than that of the other two terms. 
The subtle reason for that is in the threshold nature of the process:
the lowest possible phonon energy is $\varepsilon_{\bf k}$, not $\omega_{\rm min}$, which gives the thermal-shell
estimate $1/\tau^{(3)}\!\sim\!T^5/v^4$, much less than the result in (\ref{est_tau4}). We, therefore, give the asymptotic 
consideration only to the first two terms. 

Because the integral over the phonon energy in (\ref{4tau}) is infrared-divergent
and thus is dominated by the low-energy phonons, the asymptotic behavior of both terms is the same at low and 
high temperatures ($T\!<\!\Theta_D$ and $T\!>\!\Theta_D$). 
For $\varepsilon_{\bf k}$ being of the same order as $T$, both terms in (\ref{4tau}) yield 
an estimate of the relaxation rate 
\begin{eqnarray}
&&\frac{1}{\tau_{\bf k}^{(1)}}\approx \frac{1}{\tau_{\bf k}^{(2)}}\sim\left(\frac{g_{\rm sp}^{\rm ac}}{v}\right)^2 
\left(\frac{T\varepsilon_{\bf k}^4}{v^3 c}\right), 
\label{est_tau3}
\end{eqnarray}
which is in accord with the thermal-shell answer (\ref{est_tau4}). As we discuss below,
the  phonon-absorption term has a more complicated $\varepsilon_{\bf k}$-dependence for 
$\varepsilon_{\bf k}\!\ll\! T$.

The asymptotic result (\ref{est_tau4}) should be compared with the 
high-temperature ($T\!>\!\omega_0\!\sim\!\Theta_D$) estimate for the
optical phonon case, Eq.~(\ref{est_tau2}). The ratio of (\ref{est_tau4}) to the latter 
is $T/v$, which should imply that the contribution of the scattering on acoustic phonons is a relatively minor effect
in this temperature regime. A direct comparison is provided in Fig.~\ref{Fig1tau}, where the black line is obtained from 
(\ref{4tau}), without the use of the asymptotics. This line clearly indicates that the 
asymptotic consideration of Eqs.~(\ref{est_tau4}) and (\ref{est_tau3}) is correct and that the relaxation rate on 
thermalized acoustic phonons follows $T^5$ power law.

On a closer inspection of  Fig.~\ref{Fig1tau} we should note,  first, that the 
dominance of the acoustic phonon scattering in the low-$T$ regime  concerns a really small region 
of $T\!\alt\!\Theta_D/4$ and is unlikely to be seen in the thermal conductivity of La$_2$CuO$_4$
as this regime is known to be dominated by the grain-boundary scattering.\cite{Hess03}
Second, at the higher $T$, the relaxation rate by acoustic phonons seems to exceed the one by optical phonons, at least for 
some of the models of their DoS. This is likely to be due to  a neglect of the numerical factor difference in the 
coupling strength to acoustic and optical phonons in our effective vertices (\ref{Vac}) and (\ref{Vopt}),
and  an underestimate of the optical phonon scattering rate  by a numerical factor mentioned in Sec.~\ref{Sec_opt}.

\subsection{Summary of the phonon-scattering mechanisms}
\label{Sec_ph_sum}

Here we would like to summarize our considerations of the magnon-phonon scattering  and 
to take a broader view of its implication for the thermal conductivity.

\subsubsection{Smallness of $g$'s}

First, while the perturbative character of 
our treatment of magnon-phonon scattering is implied by the use of 
the lowest Born approximation in Fig.~\ref{diagrams} and Eq.~(\ref{1tau}), we would like to 
make it explicit that the physical range of the phenomenological magnon-phonon constants we
introduce in  (\ref{Vopt}) and (\ref{Vac}) is $g_{\rm sp}/v\!\ll\!1$. At first glance this may be surprising as the 
dependence of the  superexchange constants on the interatomic distance is often rather sharp and is 
governed by some high power of the distance, leading to estimates $\partial J/\partial a \!\approx\!\gamma J$ 
with $\gamma\!\sim\! 10-20$, see Ref.~\onlinecite{Bramwell90}. However, this largeness is offset by the
smallness of a characteristic atomic displacement associated with phonons,\cite{Bramwell90} 
$1/\sqrt{m\Theta_D}\!\sim\! 1/100$, see Appendices~\ref{appA} and \ref{appD} on how the two factors appear together 
within a microscopic approach.

\subsubsection{$k$-dependence}

Second, we note the 
importance of the $\varepsilon_{\bf k}$-dependence of the relaxation time. 
For bosons with $\varepsilon_{\bf k}\!\approx\! v |{\bf k}|$, one can estimate thermal conductivity as\cite{Herring,RC}
\begin{eqnarray}
\kappa \propto \int_0^{T/v}\tau_{\bf k} d{\bf k} \, .
\label{kappa0}
\end{eqnarray}
According to the preceding sections, in magnon-phonon $1/\tau_{\bf k}$
the lowest power   is $k^2$. In a similar situation in 1D,\cite{RC} this leads to a strong
infrared divergence of the spin component of the thermal conductivity. This means 
that the spin-phonon scattering is not sufficient to render conductivity finite and 
one needs to take into consideration other scattering mechanisms. 
In 2D, the integral in (\ref{kappa0}) still has a weak  (logarithmic) divergence for $\tau\!\sim\!k^{-2}$, 
so the (grain-)boundary scattering is sufficient to mitigate it. One of the implications of this  is
that the distribution of magnons that carry the heat most effectively is not centered at energies of order 
$T$, but is shifted toward  lower energies. More importantly, this consideration means that with the   
$\varepsilon_{\bf k}$-dependence of the relaxation rates obtained in Secs.~\ref{Sec_opt} and \ref{Sec_ac},  
magnon-phonon scattering \emph{cannot} be the only scattering mechanism and thus
must be accompanied by a boundary-like scattering in order to render magnon heat conductivity finite.

\subsubsection{Effective $1/\tau$}

Lastly, given how close the results for different models of the phonon spectra conform 
to the asymptotic expressions for $1/\tau_{\bf k}$ in (\ref{est_tau1}) and (\ref{est_tau3}), it is 
tempting to introduce a simplified, ``effective'' expression for the magnon relaxation rate on 
optical and  acoustic phonons that contains a minimal number of parameters
\begin{eqnarray}
\frac{1}{\tau_{\bf k}^{\rm eff}}\approx \sum_i a^{\rm opt}_i\,
\widetilde{\Theta}\left(T-\widetilde{\omega}_{0,i}\right)
\frac{T^2 k^2}{\omega_{0,i}} + a^{\rm ac}\,\frac{vT k^4}{\Theta_D},
\label{eff_1tau}
\end{eqnarray}
where $a^{\rm ac}\!\equiv\! (g_{\rm sp}^{\rm ac}/v)^2$ and 
$a^{\rm opt}_i\!\equiv\! (g_{{\rm sp},i}^{\rm opt}/v)^2$ are the dimensionless coupling constants to the acoustic and the 
$i$th optical mode which has the (lowest) energy $\omega_{0,i}$. The ``pseudo''-step-function $\widetilde{\Theta}(x)$
is introduced to mimic\cite{Theta} the exponential ``turn-on'' of the scattering on optical modes at the temperatures 
$T$ around $\widetilde{\omega}_{0,i}\!=\!\omega_{0,i}/2$, as in Fig.~\ref{Fig1tau}. 

We remark  that thermal conductivity obtained with this
effective $1/\tau_{\bf k}$ and appropriate set of parameters can be made virtually indistinguishable from the 
ones using  more elaborate expressions from Secs.~\ref{Sec_opt} and \ref{Sec_ac}.

\subsection{Other scattering mechanisms}
\label{Sec_others}

\subsubsection{Grain-boundary}
\label{Sec_others_grain_boundary}

As is discussed in previous section, 
boundary scattering is essential to mitigate  residual infrared
divergence of magnon $\kappa_m$ if only scattering on phonons is considered.
One can expect that in a 2D magnetic lattice even relatively weak dislocation-like defects are 
likely to act as strong boundaries for magnon propagation, similarly to the effect of crystal grain boundaries on phonons. 
Experimentally, the grain-boundary scattering is known to 
dominate entirely the low-temperature ($T\!\alt\!200$K) magnon thermal conductivity of La$_2$CuO$_4$.\cite{Hess03}
The corresponding relaxation rate is simply
\begin{eqnarray}
\frac{1}{\tau_{\bf k}^{\rm b}}\approx \frac{v}{L} ,
\label{boundary}
\end{eqnarray}
where $L$ is the characteristic size of the grain. The typical grain size quoted in Ref.~\onlinecite{Hess03} is 
$L\!\sim\!150$ lattice spacings and is in agreement with the other measurements in La$_2$CuO$_4$.

\subsubsection{Correlation length}

In the paramagnetic state above the N\'{e}el temperature, which is non-zero in most 
unfrustrated 2D AFs because of the interplane interactions and/or small anisotropies, 
finite spin-spin correlation length can be expected to represent a natural ``cut-off'' boundary for 
magnon propagation because  magnons are the spin-flips in an ordered structure.
This expectation is in agreement with a number of studies which have pointed out that 
the dynamics of the antiferromagnets is fully diffusive at scales beyond the correlation length $\xi(T)$, 
i.e., that the propagating magnons are overdamped by fluctuations at distances 
$\lambda\!\agt\!\xi(T)$.\cite{HH,CHN,Grempel,Tyc}
In effect, the correlation length acts a temperature-dependent size of an order-parameter 
domain for magnon propagation.  

The  exponential dependence of the 2D correlation length on $J/T$ has been verified experimentally in 
La$_2$CuO$_4$ and in the other 2D antiferromagnets,\cite{Kastner98,Goff,Ronnow} and is supported by extensive 
theoretical and numerical Quantum Monte-Carlo (QMC) calculations.\cite{CHN,Makivic,Hasenfratz00} 
An approximate analytical expression of it for the 
spin-$\frac12$, nearest-neighbor Heisenberg antiferromagnet on a square lattice is\cite{Greven}
\begin{eqnarray}
\xi(T)=\frac{1.13 J}{2.26J+T} \, e^{1.13J/T} ,
\label{xi}
\end{eqnarray}
in units of lattice spacings. 
While it was argued that the characteristic length-scale that defines magnon lifetime 
should also contain a temperature-dependent prefactor\cite{CHN,Grempel, Makivic,Goff}
as well as correctional factors, \cite{Hasenfratz00}
we simply suggest the following scattering rate by interpreting correlation length as a mean-free path
\begin{eqnarray}
\frac{1}{\tau_{\bf k}^{\xi}}\approx \frac{v}{\xi(T)} .
\label{tau_xi}
\end{eqnarray}
These seemingly naive expression and expectation are supported by the studies of the $T$-dependence 
of the magnon linewidth in copper-formate-tetradeuterate (CFTD), another model spin-$\frac12$ square-lattice 
antiferromagnet, by inelastic neutron scattering and QMC.\cite{Ronnow} In fact, the relaxation rate in  the form of
Eq.~(\ref{tau_xi}) has been suggested in Ref.~\onlinecite{Ronnow}, which has demonstrated that in both experiment and 
QMC results the magnon linewidth is well described  by (\ref{tau_xi}) at $T\!\agt\! 0.2J$.

Another effect of the finite correlation length is an effective gap $\Delta_\xi\!=\!v/2\xi$ 
in the magnon spectrum\cite{Takahashi}
\begin{eqnarray}
\varepsilon_{\bf k}\approx v\sqrt{k^2+\xi^{-2}(T)/4},
\label{Ek_xi}
\end{eqnarray}
which, obviously, affects the contribution of the long-wavelength excitations to thermal conductivity.

\subsubsection{Lattice disorder}

In a recent study,\cite{Ronnow14} the following scenario has been put forward. Since the ionic motion in
the cuprates  is much slower than the superexchange processes among   spins, the former
must cause a significant variation of couplings $J$ due to   zero-point and thermal lattice fluctuations. The
effect was estimated from the high-resolution neutron diffraction and 
 the zero-point motion was found responsible for the distribution of the width 
$\delta J\!\sim\!0.1J$.\cite{Ronnow14}
In a sense, this implies that magnons propagate in a medium with the velocity that is randomly varying around the 
mean value with a distribution given by $\delta J(T)$. 
One can approximate this effect by  the  $T$-dependent static random lattice disorder with
the lowest-order magnon relaxation rate due to this mechanism given by
\begin{eqnarray}
\frac{1}{\tau_{\bf k}^{\rm lat}}\approx \frac{\delta J^2(T)}{J}\, k^3 \, ,
\label{tau_imp}
\end{eqnarray}
where the temperature-dependent disorder strength can be modeled as 
$\delta J(T)\!=\!\delta J(0)\sqrt{1+4T/\Theta_D}$ to interpolate between the amplitude of zero-point and 
temperature-induced lattice fluctuations within the Debye approximation.\cite{Ziman,Ronnow14} 

Related to this mechanism are two other possible sources of magnon scattering at high temperatures 
that are harder to estimate. First is specific to La$_2$CuO$_4$, which exhibits
orthorhombic-to-tetragonal structural phase transition at about 525K
associated with a softening of a phonon mode.\cite{Kastner98} This transition may have a direct impact on the values 
of superexchange constants and also enhance magnon-phonon scattering involving the mode that is being softened.
However, it is hard to quantify both without a microscopic insight.

Second is a significant decrease and eventual collapse of $J$ at high enough temperatures, 
advocated in Ref.~\onlinecite{Bramwell90} as an ultimate result of the thermal expansion. 
With the large value of $J$ and an apparent insignificant impact of the expansion on the average value of $J$ 
in the $0-300$K range\cite{Ronnow14} in La$_2$CuO$_4$, it is hard to estimate at what 
$T$ such dramatic effects can be expected to onset.

\subsubsection{Magnon-magnon scattering}

Last  but not the least is the effect of magnon-magnon scattering.
Since these require an explicit momentum dissipation to contribute to conductivities, 
the standard Umklapp process leading to a scattering of a typical low-energy magnon with the momentum $k$ 
must involve a high-energy magnon with the energy $\varepsilon_{\rm max}\!\approx\!2.2J$. Then, one 
can suggest an ansatz
\begin{eqnarray}
\frac{1}{\tau_{\bf k}^{\rm mm,U}}\approx J k\, e^{-2.2J/T} \, .
\label{tau_mmU}
\end{eqnarray}
This can be seen as an upper limit estimate for the standard 
Umklapp scattering rate as it neglects  possible smallness  
of the matrix element and only takes into account the smallness of $k$ from the initial state.

There exists a possibility of an unconventional ``low-energy Umklapp'',
because magnons in La$_2$CuO$_4$ have two  branches,  at
${\bf k}=(0,0)$ and ${\bf k}=(\pi,\pi)$, so that some of the scattering  {\it between} these branches 
may  carry away large momenta. This logic seems to be implied in a recent calculation in 
Ref.~\onlinecite{Keimer}.  In that case, the corresponding
relaxation rate can be expected to display a power-law behavior, similar to the
``normal'' magnon-magnon scattering\cite{Chubukov}
\begin{eqnarray}
\frac{1}{\tau_{\bf k}^{\rm m-m}}\sim  
\frac{T^2 k}{v}\,, 
\label{mm_tau}
\end{eqnarray}
which, on ``thermal shell'', differs by a factor $T/\omega_0$ from the magnon-phonon 
estimate in (\ref{est_tau2}). This implies that  for sufficiently high temperatures, $T\gg\omega_0$, 
magnon-phonon scattering must be more important than the  inter-magnon scattering. 

Moreover, this expression must carry a small prefactor as it neglects the
smallnesses of the fraction of the Umklapp vs normal processes and of the phase space for scattering. 
The apparent inability of the magnon-magnon scattering theory of Ref.~\onlinecite{Keimer}
to fit the data for La$_2$CuO$_4$ from Ref.~\onlinecite{Hess03}  
beyond the boundary-controlled regime can be seen as an indirect evidence of the relative unimportance 
of this type of scattering for the magnon thermal conductivity in large-$J$ materials. We thus exclude it 
from the subsequent consideration.

\subsubsection{Comparison}

\begin{figure}[t]
\includegraphics[width=0.999\columnwidth]{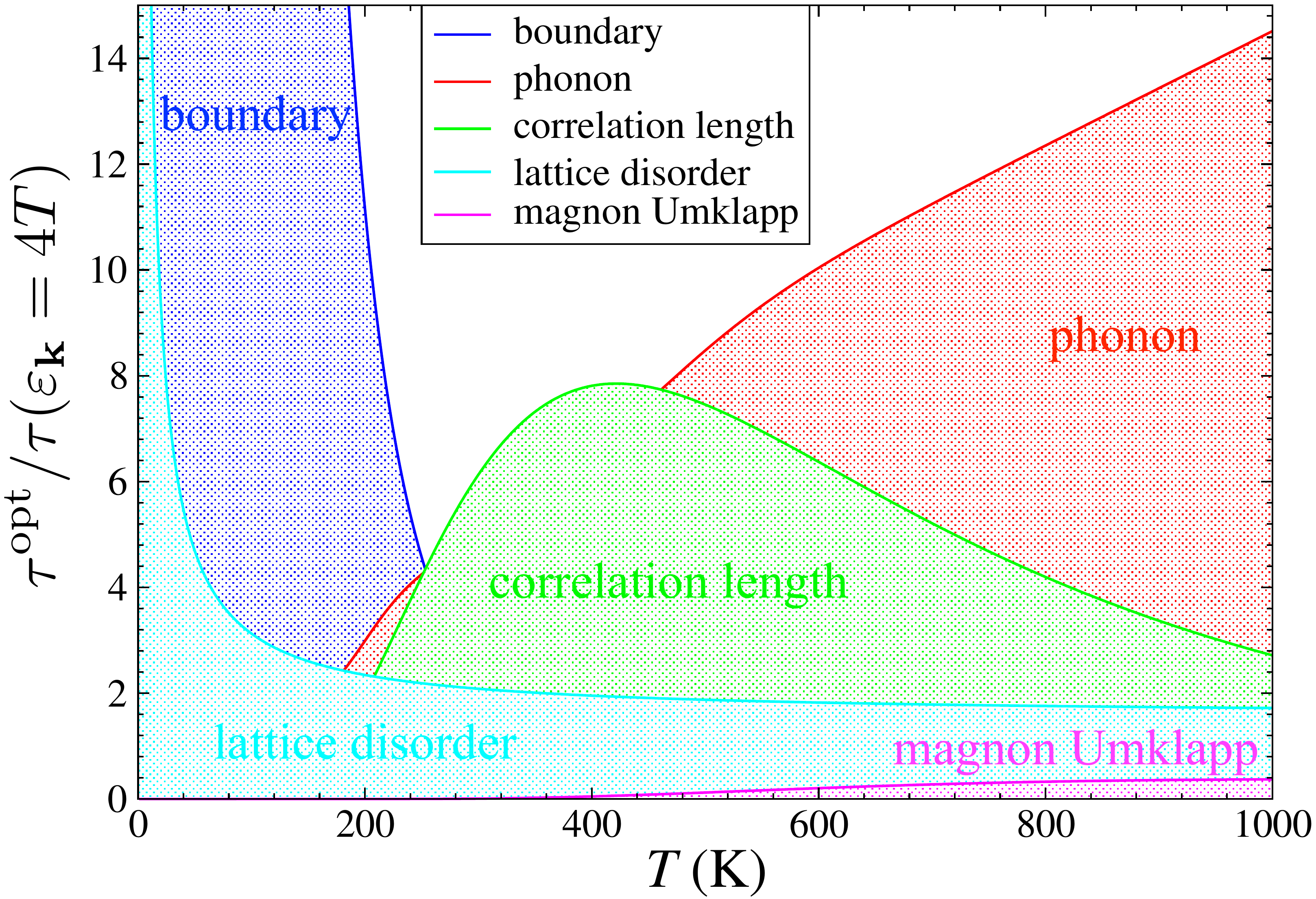}
\caption{(Color online)\ 
Magnon relaxation rates from Eqs.~(\ref{eff_1tau}), (\ref{boundary}), (\ref{tau_xi}), 
(\ref{tau_imp}), and (\ref{tau_mmU}) normalized to $T^4/v^2\Theta_D$, Eq.~(\ref{est_tau2}), at 
$\varepsilon_{\bf k}\!=\!4T$. Parameters are as described in text.}
\label{Fig_all_tau}
\vskip -0.4cm
\end{figure}

To gain a qualitative expectation for the relative contribution of the discussed scattering mechanisms
in different temperature regimes, we compile the results from Eqs.~(\ref{eff_1tau}), (\ref{boundary}), (\ref{tau_xi}), 
(\ref{tau_imp}), and (\ref{tau_mmU}) in our Fig.~\ref{Fig_all_tau}, which 
shows   relaxation rates normalized to the high-$T$ asymptote of the optical-phonon scattering in 
(\ref{est_tau2}) [$1/\tau^{\rm opt}\!=\!T^4/v^2\Theta_D$], as in Fig.~\ref{Fig1tau}. 
For the magnon-phonon scattering (\ref{eff_1tau}) we have
dimensionless coupling constants $a^{\rm opt}_1\!=\!a^{\rm ac}\!=\!0.1$ and $a^{\rm opt}_2\!=\!0.3$
with $\omega_{0,1}\!=\!\Theta_D\!=\!400$K and $\omega_{0,2}\!=\!900$K. For the boundary scattering
(\ref{boundary}) the scattering length is $L\!=\!300$ lattice spacings, and for the correlation length effect (\ref{tau_xi})
we use the expression for $\xi(T)$ in Eq.~(\ref{xi}). Temperature-dependent lattice disorder coupling in (\ref{tau_imp}) 
is chosen with $\delta J(0)\!=\!0.1J$ to roughly match the results of Ref.~\onlinecite{Ronnow14}.
$J\!=\!1500$K as before and magnon energy $\varepsilon_{\bf k}$ is chosen to be 
$\varepsilon_{\bf k}\!=\!4T$, the value of energy which roughly corresponds to the maximum of 
the thermal population of magnons at a given $T$.

One can see in Fig.~\ref{Fig_all_tau} that the scattering at low-$T$ ($T\!\alt\!\Theta_D/2$)
is controlled entirely by the grain boundaries. 
We note that the magnon-phonon scattering  and  the lattice-disorder scattering  have 
a substantial dependence on $\varepsilon_{\bf k}$ and thus lead 
to a stronger scattering   for magnons with $\varepsilon_{\bf k}\!>\!T$, but to a weaker 
scattering for $\varepsilon_{\bf k}\!<\!T$.
Thus, in Fig.~\ref{Fig_all_tau} at intermediate and high-$T$ ($T\!\agt\!\Theta_D/2$) 
the dominant contribution is due to finite correlation length and phonons.
The magnon-phonon scattering is at least as strong as 
the scattering due to correlation length at intermediate $T$ and even exceeds it  for the higher $T$
for the given choice of $\varepsilon_{\bf k}$. While
lattice disorder effect is also  significant, it remains secondary and, given its nearly perfect $T^4$ asymptotic behavior, 
may be incorporated into the magnon-phonon coupling to an optical mode. 
Thus, while the lower-energy magnons are scattered almost exclusively by the boundaries and finite correlation length,  
at $T\!\agt\!\Theta_D$ the higher-energy magnons are strongly scattered by phonons. 

Altogether, grain-boundary, correlation length, and magnon-phonon scatterings are the leading mechanisms 
of magnon energy relaxation and, therefore, have to be included in the consideration of thermal conductivity offered next.

\subsection{Thermal conductivity}
\label{Sec_kappa}

Within the Boltzmann formalism thermal conductivity by bosonic excitations in 2D is 
\begin{eqnarray}
\kappa= \sum_{\bf k} \left(\cos\varphi \, v_{\bf k}\right)^2
\left(\frac{\varepsilon_{\bf k}}{T}\right)^2\left(n_{\bf k}^2+n_{\bf k}\right)\, \tau_{\bf k}\, ,
\label{kappa}
\end{eqnarray}
where $\varphi$ is the angle between ${\bf v}_{\bf k}$ and the current. Using the linearized form of magnon 
dispersion (\ref{Ek}) and the fact that all of the discussed relaxation rates are isotropic in ${\bf k}$ 
we can simplify (\ref{kappa}) to a 1D integral
\begin{eqnarray}
\kappa\approx \frac{T^2}{2\pi}\int_0^{x_{\rm max}} dx \, \frac{x^3\, e^x}{(e^x-1)^2}
\,\tau(x,T)\, ,
\label{kappa1}
\end{eqnarray}
where $x\!=\!\varepsilon_{\bf k}/T$ and $x_{\rm max}\!=\!\varepsilon_{\rm max}/T$ with 
$\varepsilon_{\rm max}\!=\!v\sqrt{2\pi}$, in which we used the Debye-like approximation for magnons and
also accounted for two magnon modes in the Brillouin zone. 

At finite correlation length $\xi$, the magnon spectrum opens a gap\cite{Takahashi} 
according to (\ref{Ek_xi}), which results in a modification of the expression for the thermal conductivity 
\begin{eqnarray}
\kappa\approx \frac{T^2}{2\pi}\int_{x_{\rm min}}^{x_{\rm max}}  dx \left(1-\frac{x^2_{\rm min}}{x^2}\right)
\, \frac{x^3\, e^x}{(e^x-1)^2}\,\tau(x,T)\, ,
\label{kappa_xi}
\end{eqnarray}
where $x_{\rm min}\!=\!\Delta_\xi/T$ with $\Delta_\xi\!=\!v/2\xi(T)$.

\subsubsection{Comparison}
\label{RefCompar}
\begin{figure}[t]
\includegraphics[width=0.999\columnwidth]{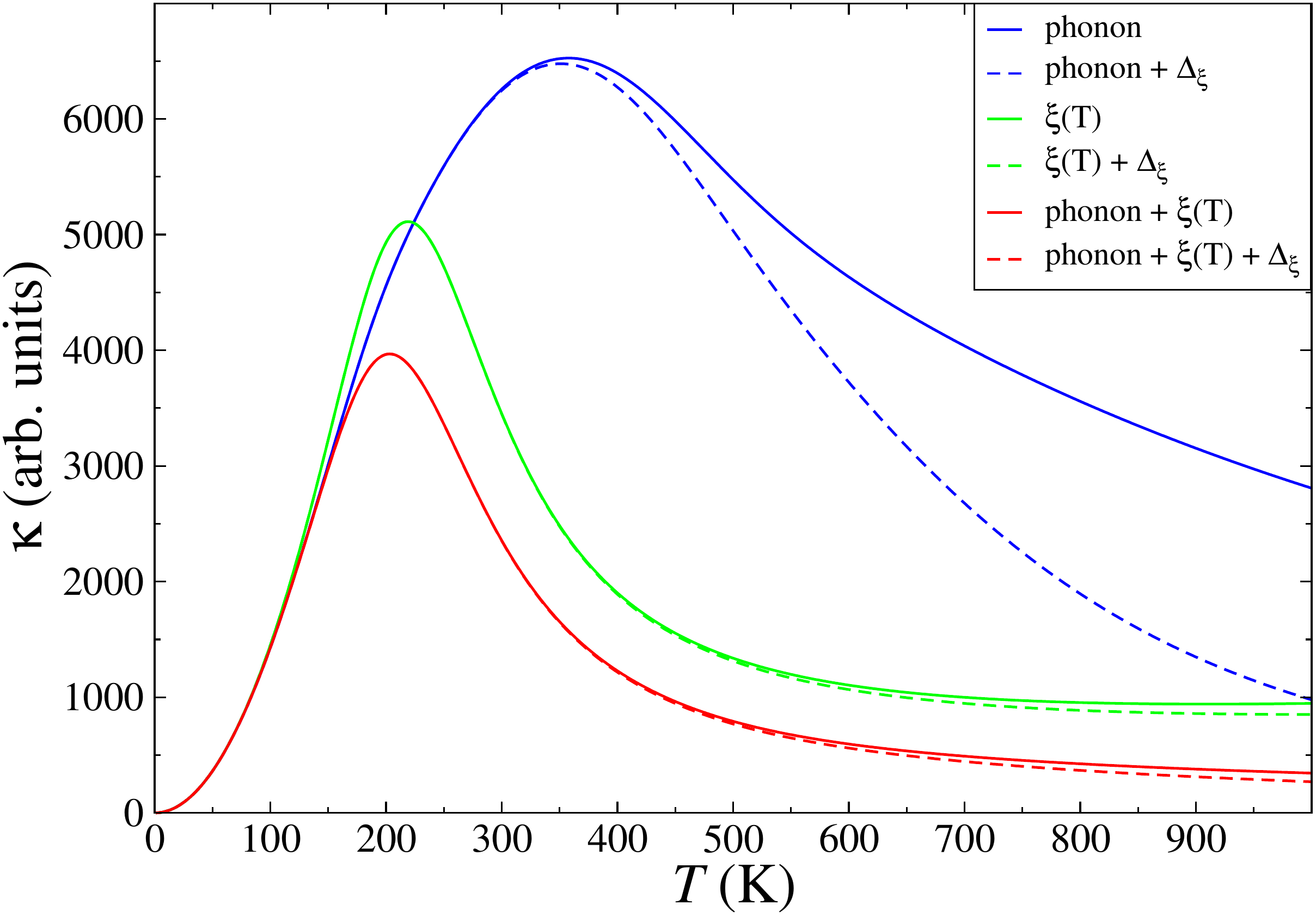}
\caption{(Color online)\ Magnon $\kappa(T)$ for the scattering by the grain boundaries, $L\!=\!300$, and either the 
correlation length (middle curves), Eq.~(\ref{tau_xi}), or phonon scattering (upper curves), Eq.~(\ref{eff_1tau}). 
Lower curves combine the effects of all three, grain-boundary, phonon, and correlation length scattering mechanisms.
Parameters for the phonon scattering are as in Fig.~\ref{Fig_all_tau}. 
Dashed lines show the effect of the gap in  magnon spectrum (\ref{kappa_xi}).}
\label{Fig_kappa}
\vskip -0.4cm
\end{figure} 

\begin{figure}[t]
\includegraphics[width=0.999\columnwidth]{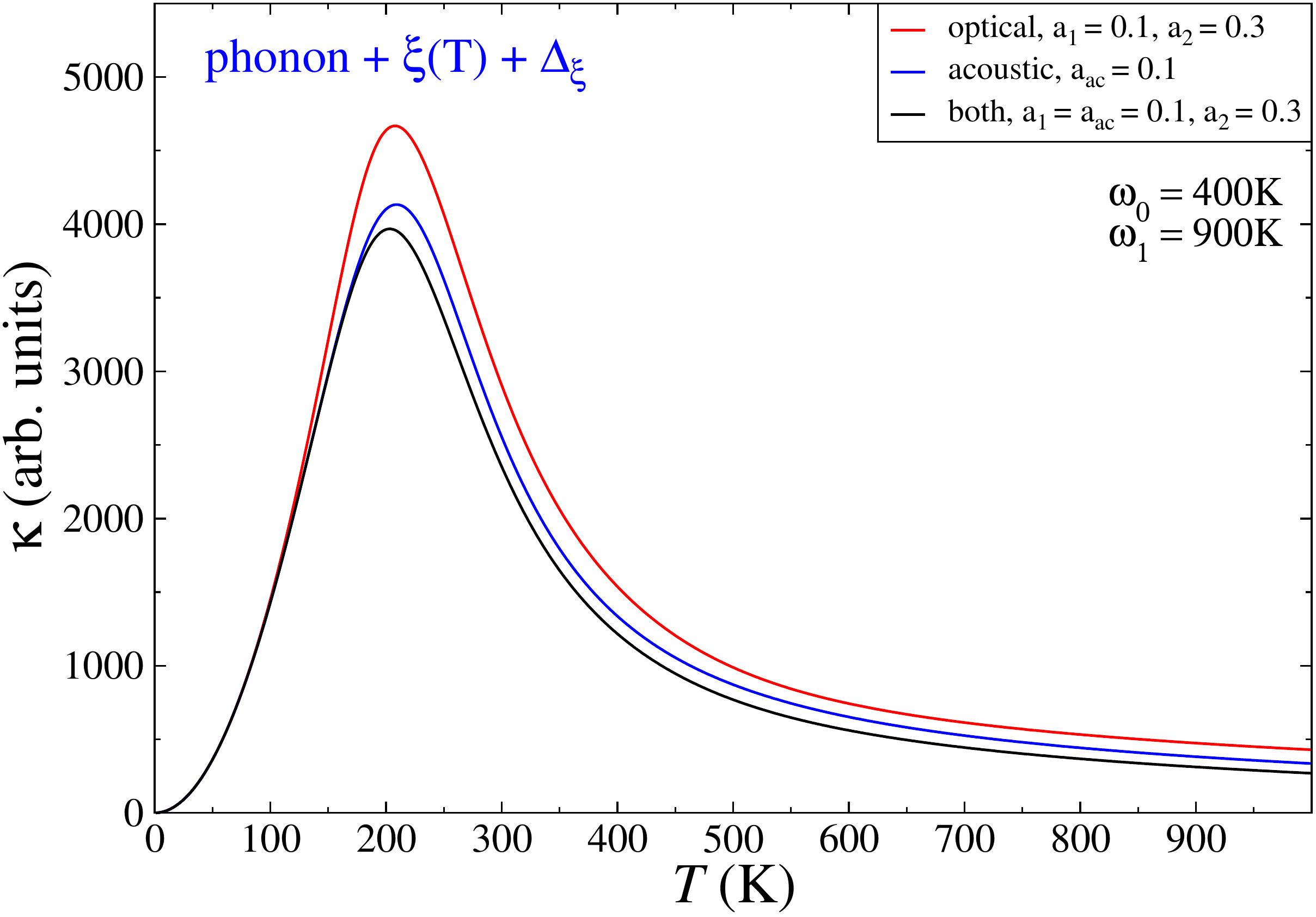}
\caption{(Color online)\ Magnon $\kappa(T)$ with all three key scatterings, same as in Fig.~\ref{Fig_kappa}.
For the magnon-phonon scattering, results for coupling to only optical, only acoustic, and to both are shown.}
\label{Fig_kappa2}
\vskip -0.4cm
\end{figure}

Finally, we present the results of our calculations of magnon thermal conductivity in 
Figs.~\ref{Fig_kappa} and \ref{Fig_kappa2}, in which we demonstrate the effects of 
 boundary scattering (\ref{boundary}) together with the  finite correlation length and/or magnon-phonon
relaxation mechanisms. In Fig.~\ref{Fig_kappa}, the upper solid curve 
shows $\kappa(T)$ for the case when the scattering is only by the grain boundaries (\ref{boundary})
with $L\!=\!300$ and by phonons (\ref{eff_1tau}) with the same coupling parameters 
as in Fig.~\ref{Fig_all_tau}. The middle solid curve is for $\kappa(T)$ due to grain boundaries 
(\ref{boundary}) and
the correlation length (\ref{tau_xi}). The corresponding dashed lines are for the same cases, but with an additional 
effect of the finite-$T$ gap in the magnon spectrum  $\Delta_\xi$ due to finite correlation length, Eq.~(\ref{kappa_xi}).
The effect is minimal on the middle curve, but is rather dramatic in the phonon-scattering case. This 
is due to strong $\varepsilon_{\bf k}$-dependence of the magnon-phonon scattering discussed in Sec.~\ref{Sec_others}, 
which 
changes substantially the magnon population contributing to the heat current. Namely, in the ``boundary+$\xi(T)$'' case,
the typical magnon in (\ref{kappa_xi}) is a thermalized one, $\varepsilon_{\bf k}\!\sim\!T$,
so the exponentially small gap $\Delta_\xi(T)$ of (\ref{Ek_xi}) does not affect it. In the ``boundary+phonon'' case,
the high-energy magnons are scattered strongly by phonons, see Fig.~\ref{Fig_all_tau}, while the heat-conducting
population of magnons leans strongly to the low energies, hence a dramatic impact of opening the gap.

Lower curves shows the result of combining all three,  the
grain-boundary, the phonon, and the correlation length scattering mechanisms.
Clearly, the phonon relaxation mechanism leads to a stronger scattering 
at the higher-$\varepsilon_{\bf k}$ part of the heat-conducting magnon population, reducing the
overall conductivity, while the longer-wavelength
part is now controlled by the correlation length, evidenced by a weak sensitivity of $\kappa(T)$ to the gap in the magnon 
spectrum (dashed curves).

Fig.~\ref{Fig_kappa2} complements this study with the consideration of 
three different scenarios: the magnon-phonon coupling  is to (i) two optical modes, one at 
$\omega_{0,1}\!=\!400$K and the other at $\omega_{0,2}\!=\!900$K (an analog of the high-energy stretching mode), 
(ii) to an acoustic branch of phonons only, and (iii) to both (same curve as in Fig.~\ref{Fig_kappa}). 
One can see that the overall effect on $\kappa(T)$ is very similar in all three cases. This is natural
in the light of the preceding discussion that points to the significant scattering effect  by phonons at 
higher energies, which can be expected to be comparable regardless of the nature of the phonon branches.
 We would also like to point out that we have performed the same calculation of $\kappa(T)$ in
(\ref{kappa_xi}) using not the ``effective'' magnon-phonon relaxation rate, proposed in Sec.~\ref{Sec_ph_sum} in 
Eq.~(\ref{eff_1tau}), but the relaxation rates in Secs.~\ref{Sec_opt} and \ref{Sec_ac}, Eqs.~(\ref{3tau}) and 
(\ref{4tau}), using different effective phonon DoS models, previously matched to a direct numerical integration in 
Eq.~(\ref{1tau}). Aside for the adjustments of the magnon-phonon coupling constants needed in the cases when  
phonon spectral weight is spread over a substantial energy range, these calculations have produced the results that 
are virtually identical to the ones in Figs.~\ref{Fig_kappa}  and \ref{Fig_kappa2}.

Altogether, magnon thermal conductivity in La$_2$CuO$_4$ should be  largely controlled by the ``boundary-like''
scatterings, coming from either the real grain-boundaries or the 2D correlation length, with a substantial correction 
from the magnon-phonon scattering, affecting $\kappa(T)$ at intermediate and high temperatures. 

\subsubsection{Remarks}

For the latter regime, there are two additional remarks that need to be made.
First, according to Eq.~(\ref{xi}), at 800K ($\approx\!0.5J$) the correlation length is of  order of 3 lattice spacings.
Is the magnon picture still applicable and is it realistic to expect further reduction of the mean-free path by a 
phonon scattering? We believe that although the magnon description at $T\!\agt\!0.5J$ may be an extrapolation,
such a correlation length corresponds to a ``patch'' of about 30 spins, which is known to be well-described
by the spin-waves in finite-cluster studies.  A somewhat more relevant quantity is the effective
gap in the magnon spectrum at this temperature, $\Delta_\xi\!\approx\!J/4$, which is still considerably smaller than the magnon 
bandwidth ($2J$). Given that the magnon-phonon scattering mostly affects magnons with 
$\varepsilon_{\bf k}\!>\!T$, it seems perfectly legitimate that the phonons \emph{are} the source of a further shortening 
of the mean-free path in this regime.

Second, the use of the bosonic description of the thermodynamics of spin excitations of the $S\!=\!\frac12$
Heisenberg model on a square lattice is limited by the temperatures of order $0.6J$, at which the specific 
heat shows a broad maximum\cite{Sandvik_Cv} while within the bosonic description specific heat 
saturates at a somewhat higher $T$.
For a comparison of our results with experimental data this implies that the high-temperature tail of our $\kappa(T)$ 
should only serve as an upper limit estimate of the reality.

\section{Memory function approach}
\label{MF}

We now turn to the memory function approach, which does not proceed via one
particle excitations, but focuses on the dynamics of the current directly,
as we will sketch next. We start from the magnon heat current
in terms of Bogoliubov quasiparticles
\begin{equation}
\mathbf{j}=\sum_{\mathbf{k}}\varepsilon_{\mathbf{k}}\mathbf{v}_{\mathbf{k}}
\beta_{\mathbf{k}}^{\dagger}\beta_{\mathbf{k}}^{\vphantom{\dagger}}
\,,\label{10}
\end{equation}
where $\mathbf{v}_{\mathbf{k}}=\nabla_{\mathbf{k}}\varepsilon_{\mathbf{k}}$
is the velocity of the magnon. Following the memory function method
\cite{Forster1975}, the dynamical thermal conductivity tensor at frequency
$z=\omega+i0^{+}$ is
\begin{equation}
VT\kappa_{\mu\nu}{=}\left\langle j_{\mu}\left|\frac{i}
{z-{\cal L}}\right.j_{\nu}
\right\rangle {=}\left(\frac{i}{z-\mathbf{M}(z){\boldsymbol{\chi}}^{-1}}
\boldsymbol{\chi}\right)_{\mu\nu}
,\label{11}
\end{equation}
where $V$ is the volume, and ${\cal L} ={\cal L}_{{\rm s}} +{\cal L}_{{\rm
ph}} +{\cal L}_{{\rm s-ph}}={\cal L}_{0}+{\cal L}_{{\rm s-ph}}$ is the
Liouville operator ${\cal L}A=[{\cal H},A]$, comprising spin, phonon, and
spin-phonon parts, see (\ref{0}). $\langle A|B\rangle= \int_{0}^{\beta}
\langle A^{+}(\lambda)B\rangle d\lambda-\beta\langle A^{+}\rangle \langle
B\rangle$ is the Mori's scalar product with $A(\lambda)= e^{\lambda{\cal
H}} A e^{- \lambda{\cal H}}$ and $\beta=1/T$ is the inverse
temperature. $\boldsymbol{\chi}$ is the isothermal heat current
susceptibility, $\chi_{\mu\nu}= \langle j_{\mu}| j_{\nu}\rangle$, and
$M_{\mu\nu}(z)= \langle{\cal L} j_{\mu}| (z-Q{\cal L})^{-1} Q{\cal
L}j_{\nu} \rangle$ is the memory function matrix. $Q$ is a projector
perpendicular to the heat current in terms of Mori's product $Q=1-
\sum_{\mu\nu}| j_{\mu}\rangle \chi_{\mu\nu}^{-1}\langle j_{\nu}|$.

We will evaluate (\ref{11}) for $\omega\rightarrow0$ to order $O({\cal H}_{
{\rm s-ph}}^{2})$.  Within perturbation theory \cite{Forster1975} to the
leading order in the spin-phonon coupling, the memory matrix is given by
\begin{equation}
M_{\mu\nu}(z)=\left\langle [{\cal H}_{{\rm s-ph}},j_{\mu}]\left|
\frac{1}{z-{\cal L}_{0}}\right.[{\cal H}_{{\rm s-ph}},j_{\nu}]
\right\rangle _{0}
\,,\label{12}
\end{equation}
where the subscript $0$ refers to Mori's product and thermal averaging
with respect to the canonical ensemble of the system at zero coupling,
${\cal H}_{{\rm s-ph}}=0$. This subscript will be dropped hereafter.
Since $M_{\mu\nu}\sim O({\cal H}_{{\rm s-ph}}^{2})$ already, the
static current susceptibility is needed only for ${\cal H}_{{\rm s-ph}}=0$.
With $[{\cal H}_{0},j_{\mu}]=0$ and $\langle j_{\mu}\rangle=0$ 
\begin{equation}
\chi_{\mu\nu}=\frac{\delta_{\mu\nu}}{T}\langle j_{x}j_{x}
\rangle=\frac{\delta_{\mu\nu}}{2T}
\sum_{\mathbf{k}}\varepsilon_{\mathbf{k}}^{2}
\mathbf{v}_{\mathbf{k}}^{2}n(\varepsilon_{\mathbf{k}})
\left(1+n(\varepsilon_{\mathbf{k}})\right)
\,,\label{13}
\end{equation}
where $n(\varepsilon_{\mathbf{k}})=1/(e^{\varepsilon_{ \mathbf{k}}/T}-1)$
is the Bose-function.

\subsection{Relaxation time approximation}
\label{sub:Relaxation-time-approximation}

The relaxation-time approximation corresponds to replacing
$\mathbf{M}(z)/\boldsymbol{\chi}$ by $-i/\tau$ with a phenomenological
scattering time $\tau$. In that case (\ref{11}) reads
\begin{equation}
\kappa=\frac{1}{2T^{2}}\sum_{\mathbf{k}}\varepsilon_{\mathbf{k}}^{2}
\mathbf{v}_{\mathbf{k}}^{2}n(\varepsilon_{\mathbf{k}})
\left(1+n(\varepsilon_{\mathbf{k}})\right)\,\tau
\,,\label{14}
\end{equation}
with $\kappa_{\mu\nu}=\delta_{\mu\nu}\kappa$. This is identical to the
standard result from kinetic theory. For $T\ll J$, Eq. (\ref{14}) yields
$\kappa\propto T^{2}$, as expected for bosons in $2D$.  While the
case $T\gg J$ is unphysical, we note that then $\kappa\propto
const$.

We note that (\ref{14}) is completely identical to (\ref{kappa}) for a
momentum-independent scattering rate. In principle, the memory function
approach in Eq. (\ref{11}) can be formulated with a particle-hole type of
observable, $\mathbf{j}_{\mathbf{kq}}= \varepsilon_{\mathbf{k}}
\mathbf{v}_{\mathbf{k}} \beta_{\mathbf{k}}^{\dagger} \beta_{\mathbf{k}+
\mathbf{q}}^{\vphantom{\dagger}}$, in order to explicitly analyze momentum-dependent 
current relaxations rates. We will not pursue this direction here.

\subsection{Spin-phonon relaxation rate}

Here we would like to restrict ourselves to the case of acoustic phonons.
This reduces the problem to a monoatomic Bravais lattice with ionic
masses $m$ where magnetoelastic couplings results from stretching
of the nearest-neighbor exchange bonds along the in-plane bond directions,
the case considered in detail in Appendix~\ref{appA}. Lattice deformations
perpendicular to the bond direction lead to higher-order couplings.

First, we use explicit form of ${\cal H}_{{\rm s-ph}}$ in (\ref{Hsph2_0}) to
find the commutator $[{\cal H}_{{\rm s-ph}},j_{\mu}]$, which has a meaning
of a force
\begin{equation}
[{\cal H}_{{\rm s-ph}},j_{\mu}]=\frac{1}{2N}
\sum_{\mathbf{k}{\mathbf{q}},\ell}
B_{\mathbf{k'}}^{\dagger}\,
\mathbf{F}_{\mathbf{k},\mathbf{q}}^{\ell\mu}\,
B_{\mathbf{k}}^{\vphantom{\dagger}}\,
\left(a_{-{\mathbf{q}}\ell}^{\dagger}+
a_{{\mathbf{q}}\ell}^{\vphantom{\dagger}}\right)
\,,\label{15}
\end{equation}
where ${\bf k'}= {\bf k}+ {\bf q}_{\parallel}$ imposed by the conservation
of the in-plane component of the momentum, $\ell$ numerates acoustic phonon
branches, $B_{\mathbf{k}}^{\dagger}= (\beta_{\mathbf{k}}^{\dagger},
\beta_{\mathbf{k}}^{\vphantom{\dagger}})$, and the matrix
\begin{eqnarray}
\mathbf{F}_{\mathbf{k},\mathbf{q}}^{\ell\mu} & = & 
\left(e_{\mathbf{k}}^{\mu}-e_{\mathbf{k}+ 
\mathbf{q}_{\parallel}}^{\mu}\right)
\mathbf{V}_{\mathbf{k},{\bf k}+{\bf q}_{\parallel},{\mathbf{q}}}^{\ell}
\,,\label{16}
\end{eqnarray}
where we have introduced the shorthand notation for the energy current of a
single magnon mode $\mathbf{e}_{\mathbf{k}}= \varepsilon_{\mathbf{k}}
\mathbf{v}_{\mathbf{k}}$ and used inversion symmetry
$\varepsilon_{-\mathbf{k}}= \varepsilon_{\mathbf{k}}$,
$\mathbf{v}_{-\mathbf{k}}= -\mathbf{v}_{\mathbf{k}}$. The $2\times2$
``vertex matrix'' $\mathbf{V}_{\mathbf{k},{\bf k'},{\mathbf{q}}}^{\ell}$ is
built from the diagonal elements responsible for the ``normal'' scattering
processes, $V_{{\bf k,k',q}}^{\ell}$ given in (\ref{Vmagph}), and the
off-diagonal, ``anomalous'' terms $V_{{\bf k,k',q}}^{{\rm od},\ell}$ given
in (\ref{Vmagph1}).

As usual \cite{Forster1975} (\ref{12}) can be rewritten as
$z\,M_{\mu\nu}(z)= \chi_{\mu\nu}^{F}(z)- \chi_{\mu\nu}^{F}$, where
$\chi_{\mu\nu}^{F}(z)$ is the retarded \emph{dynamical} force-force
susceptibility, resulting from analytic continuation of the imaginary-time
Greens function $\chi_{\mu\nu}^{F}(\tau)= \langle T_{\tau}([{\cal H}_{{\rm
s-ph}}, j_{\mu}]^{+}(\tau)[{\cal H}_{{\rm s-ph}},j_{\mu}])\rangle$, and
$\chi_{\mu\nu}^{F}=\langle[{\cal H}_{{\rm s-ph}},j_{\mu}]| [{\cal H}_{{\rm
s-ph}}, j_{\mu}]\rangle$ is the \emph{isothermal} force-force
susceptibility. Due to tetragonal symmetry, all of these quantities
are diagonal with respect to $\mu,\nu$.

\begin{figure}[tb]
\begin{centering}
\includegraphics[width=0.7\columnwidth]{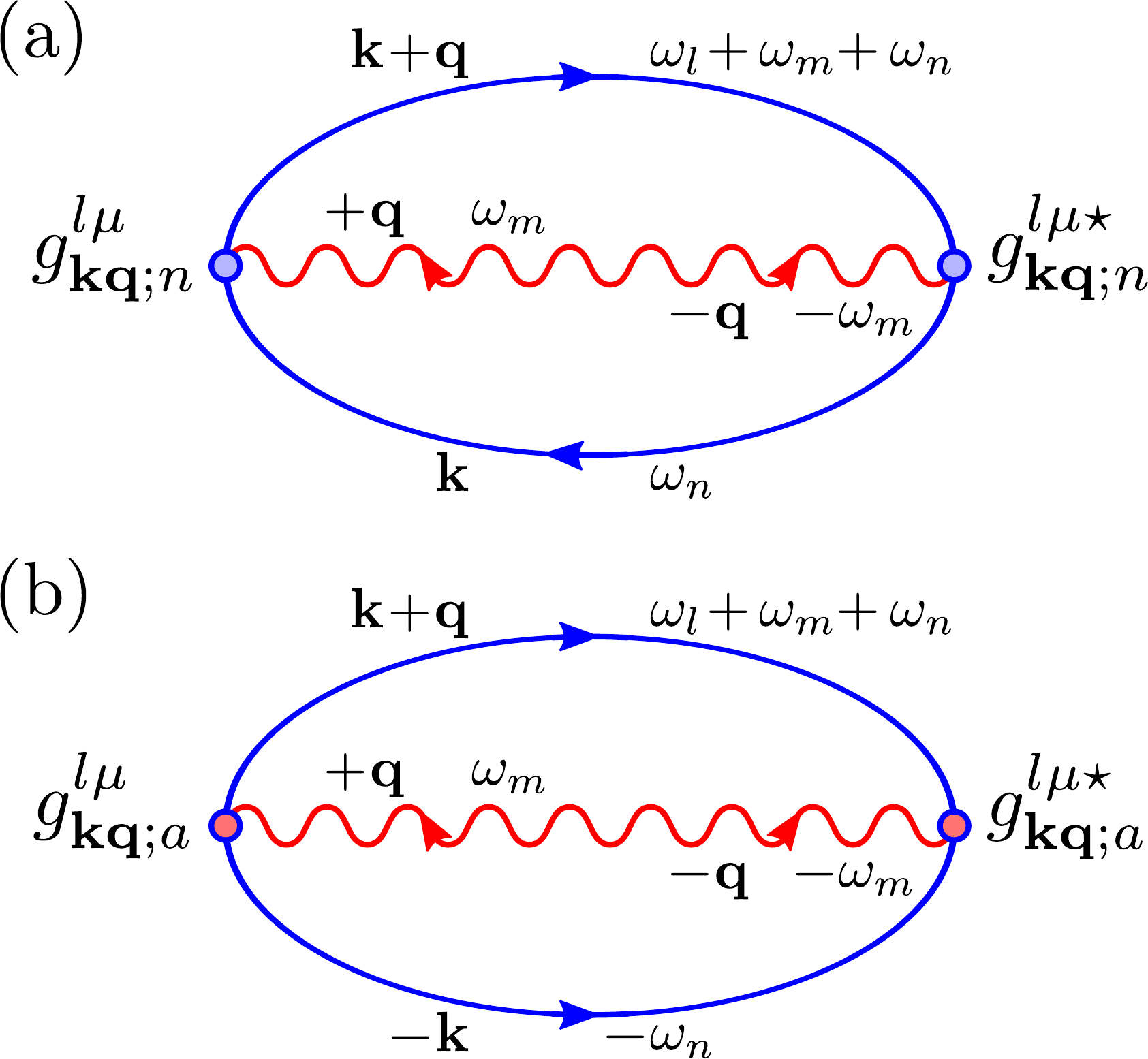} 
\par\end{centering}

\protect\protect\caption{\label{fig1}(Color online)\ Force-force
susceptibility diagrams. Solid lines are magnons, wavy lines are phonons.
(a) normal contribution (magnon emission\--absorption), (b) anomalous
contribution (two magnon emission (absorption)). Filled
circles in (a) are normal and in (b) anomalous force\--vertices.}
\end{figure}

For thermal transport we focus on the DC limit $\omega{\rightarrow}0$.
Since $\mathrm{Re}[M_{\mu\nu}(i0^{+})]=0$ and $\mathrm{Im}\chi_{\mu\nu}^{F}=0$,
we have to calculate $\mathrm{Im}[\chi_{\mu\nu}^{F}(z=i0^{+})]$ only.
This can be done using the two diagrams in Fig. \ref{fig1}, which
comprise \emph{normal} magnon emission-absorption and \emph{anomalous}
two-magnon emission (absorption) processes, i.e. $\chi_{\mu\nu}^{Fn}(z)$
and $\chi_{\mu\nu}^{Fa_{1,2}}(z)$. After some algebra we get 
\begin{eqnarray}
\lefteqn{M_{\mu\nu}(i0^{+})=-i\,
\frac{\pi\delta_{\mu\nu}}{T}\sum_{\mathbf{k},{\mathbf{q}},\ell}
n(\omega_{\mathbf{q}}^{\ell})}
\label{17}\\
 &  & \times\Big(2|g_{\mathbf{k}{\mathbf{q}};n}^{\ell\mu}|^{2}
n(\varepsilon_{\mathbf{k}})\left(1{+}n(\varepsilon_{\mathbf{k}+
\mathbf{q}_{\parallel}})\right)\delta\left(\varepsilon_{\mathbf{k}}{-}
\varepsilon_{\mathbf{k}+\mathbf{q}_{\parallel}}{+}\omega_{\mathbf{q}}^{\ell}\right)
\nonumber \\
 &  & +|g_{\mathbf{k}{\mathbf{q}};a}^{\ell\mu}|^{2}\big(1{+}
n(\varepsilon_{\mathbf{k}})\big)\left(1{+}n(\varepsilon_{\mathbf{k}+
\mathbf{q}_{\parallel}})\right)\delta\left(\varepsilon_{\mathbf{k}}{+}
\varepsilon_{\mathbf{k}+\mathbf{q}_{\parallel}}{+}\omega_{\mathbf{q}}^{\ell}
\right)\Big),
\nonumber 
\end{eqnarray}
where the ``force vertices'' $g_{\mathbf{k}{\mathbf{q}};n(a)}^{\ell\mu}$
refer to the normal (subscript $n$) and anomalous (subscript $a$) processes
\begin{eqnarray}
 &  & g_{\mathbf{k}{\mathbf{q}};n}^{\ell\mu}=
\left(e_{\mathbf{k}}^{\mu}-e_{\mathbf{k}+\mathbf{q}_{\parallel}}^{\mu}\right)\,
V_{{\bf k},{\bf k}+{\bf q}_{\parallel},{\bf q}}^{\ell},
\label{18}\\
 &  & g_{\mathbf{k}{\mathbf{q}};a}^{\ell\mu}=
\left(e_{\mathbf{k}}^{\mu}-e_{\mathbf{k}+\mathbf{q}_{\parallel}}^{\mu}\right)\,
V_{{\bf k},{\bf k}+{\bf q}_{\parallel},{\bf q}}^{{\rm od,\ell}},
\end{eqnarray}
with the explicit expressions for vertices given in (\ref{Vmagph}) and
(\ref{Vmagph1}). We note that $\mathrm{Im} [M_{\mu\mu}^{n(a)}(i0^{+})]
\leqslant0$ separately for both normal and anomalous contributions, as to
be expected from causality.

\subsection{Phonon dispersion}

In the following sections we provide a discussion of several aspects of
thermal conductivity obtained from the memory function approach.  We are
going to discuss orders of magnitude estimates for the cuprates, the
asymptotic behavior of the scattering rate, and the results from the
numerical evaluation of the memory function. Whenever of interest, explicit
reference will be made to parameters relevant to La$_{2}$CuO$_{4}$.  In
what follows, we parametrize phonon excitations that are relevant for
spin-phonon coupling as given by two degenerate modes that are polarized
in-plane and have the dispersion
\begin{equation}
\omega^\ell_{{\mathbf{q}}}=
\Theta_{D}\,\sqrt{\sum_{\alpha}\sin\left(\frac{q_{\alpha}}{2}\right)^{2}}
\,.\label{20}
\end{equation}
$\Theta_{D}$ is the Debye energy for phonons as before. Since we set
$\hbar=k_{B}=1$ and also assume all lengths in units of lattice constant
unless mentioned otherwise, $\Theta_{D}/2$ and the sound velocity $c$ will
be used interchangeably.

\subsection{Asymptotic scattering rates}
\label{Sec:asymptotes}

For the ``normal'' emission-absorption of acoustic phonons in
Fig.~\ref{fig1}(a) we expect four asymptotic temperature regimes, which are
determined in part by the kinematics of the magnon-phonon scattering. (i)
For $T\ll\Theta_{D}$, thermalized magnon-phonon scattering involves only
long-wavelength excitations. (ii) As $T$ is increased, scattering from the
zone-boundary, optical-like phonons with $\omega_{{\mathbf{q}}}\lesssim T$
and a high density of states begin contribute with the Arrhenius-like,
activated $T$-dependence. For the dispersion in (\ref{20}) such phonons
occur at the corners of the BZ at $\mathbf{q} \sim\mathbf{Q}_{AF}=
(\pi,\pi)$.  (iii) For $\omega_{\mathbf{Q}_{AF},q_{\perp}}\ll T\ll J$,
magnons can be scattered by thermalized phonons from $\mathbf{k} \!=
\!(0,0)$ to $\mathbf{Q}_{AF}$. (iv) Finally, one can consider a regime
$T\gg J$ in (\ref{17}), which however is unphysical regarding the
definition of magnons.

\subsubsection{Acoustic phonon regime}

First, we consider the ``normal'' scattering processes and
$T\ll\Theta_{D}$.  For that we may use the long-wavelength limit
$|\mathbf{k}+ \mathbf{q}_{\parallel}|, |{\bf k}|,|{\bf q}_{\parallel} |\!
\ll\!1$ for all the in-plane wave vectors. As discussed in detail in
Appendix~\ref{appA}, the in-plane modulation of $J$ will only couple to the
two out of three phonon branches, whose in-plane polarization can be always
chosen as $|${\boldmath$\xi$}$_{{\bf q}}^{\ell}|\!\approx\!1$. We also
approximate $\omega_{{\mathbf{q}}}^{\ell}\! \approx\! c|{\bf q}|\!= \! c
\sqrt{|{\bf q}_{\parallel} |^{2}+|{\bf q}_{\perp}|^{2}}$, with a sound
velocity much less then the magnon velocity, i.e. $c\ll v$.  Expanding in
(\ref{18}) to lowest order in $|\mathbf{k}+ \mathbf{q}_{\parallel}|$,
$|{\bf k}|$, and $|{\bf q}_{\parallel}|$, the memory function reads
\begin{eqnarray}
\lefteqn{M_{\mu\nu}(i0^{+})=-i\,\delta_{\mu\nu}\frac{2\pi
\left(g_{{\rm sp}}^{{\rm ac}}\right)^{2}}{T}}
\label{22}\\
 &  & \hphantom{M_{\mu\nu}(i0^{+})}\times
\sum_{\mathbf{k},\mathbf{q},\ell}
\frac{|{\bf q}_{\parallel}|^{2}|{\bf k}||
\mathbf{k}+\mathbf{q}_{\parallel}|\left(e_{\mathbf{k}}^{\mu}-
e_{\mathbf{k}+\mathbf{q}_{\parallel}}^{\mu}\right)^{2}}
{\sqrt{|{\bf q}_{\parallel}|^{2}+|{\bf q}_{\perp}|^{2}}}
\nonumber \\
 &  & \hphantom{aa}\times n(\omega_{\mathbf{q}}^{\ell})
n(\varepsilon_{\mathbf{k}})\left(1{+}n(\varepsilon_{\mathbf{k}+
\mathbf{q}_{\parallel}})\right)
\delta\left(\varepsilon_{\mathbf{k}}{-}\varepsilon_{\mathbf{k}+
\mathbf{q}_{\parallel}}{+}\omega_{\mathbf{q}}^{\ell}\right),
\nonumber 
\end{eqnarray}
where $g_{{\rm sp}}^{{\rm ac}}\!=\!S \lambda/ (2\sqrt{mc})$ with $\lambda
\!= \!\partial J/ \partial r$, see Appendix~\ref{appA} and
Eq. (\ref{Vmagph3}). First, the magnon and phonon distribution functions
$n( \varepsilon_{ \mathbf{k}})$ and $n( \omega_{ \mathbf{q}}^{\ell})$ in
(\ref{22}) imply that $|{\bf k}| \! \sim\!T/v$ and $|{\bf
q}|\!\sim\!T/c$. From this and using energy conservation $\varepsilon_{
\mathbf{k}+ \mathbf{q}_{ \parallel}} \! = \! \varepsilon_{ \mathbf{k}} \!
+\! \omega_{\mathbf{q}}^{\ell}$, we conclude that the in-plane phonon
momentum is of magnitude $|\mathbf{q}_{\parallel}| \! \sim \! T/v$,
i.e. the dominant contribution of the phonon momentum is
\emph{out-of-plane}.  This coincides with the conclusion reached in
Sec.~\ref{prelim} after Eq.~(\ref{acoustic}). In turn, $|\mathbf{k}+
\mathbf{q}_{ \parallel} | \! \sim \! T/v$ and $(e_{ \mathbf{k}}^{\mu}- e_{
\mathbf{k}+ \mathbf{q}_{ \parallel}}^{ \mu})^{2} \! \sim \! v^{2}T^{2}$.
Then, a na\"{i}ve power-counting, accounting also for factors of magnon and
phonon velocities, $v$ and $c$, and keeping in mind that the
$\delta(\ldots)$-function in (\ref{22}) replaces one $k$-integration by a
factor of $1/v$, suggests that $M \! \sim \!( g_{{\rm sp}}^{ {\rm ac}})^{
2}T^{ 8}/v^{ 6}$.  However, this approach misses one subtle detail,
i.e. that the $q_{\perp}$-integration is singular with
\begin{equation}
\int_{0}^{T/c}dq_{\perp} \frac{n(\omega_{\mathbf{q}}^{\ell})}
{\sqrt{|{\bf q}_{\parallel}|^{2}+ |{\bf q}_{\perp}|^{2}}}
\sim\int_{T/v}^{T/c}dq_{\perp}\frac{T}{cq_{\perp}^{2}}\sim\frac{v}{c}
\,,\label{23}
\end{equation}
which introduces an additional factor of $(v/c)$ and finally leads to
\begin{equation}
M_{\mu\nu}^{n}(i0^{+}) \propto -i\, \delta_{\mu\nu}
\left(g_{{\rm sp}}^{{\rm ac}}\right)^{2}\,\frac{T^{8}}{cv^{5}}
\,.\label{24}
\end{equation}
From (\ref{13}) we have $\chi_{\mu\nu}\sim T^{3}$ with a dimensionless
prefactor of order unity. Therefore, using $c\sim\Theta_{D}$, the
scattering rate is
\begin{equation}
\frac{1}{\tau_{ac}^{n}}\propto\frac{T^{5}}{\Theta_{D}v^{3}}
\,,\label{25}
\end{equation}
with the dimensionless prefactor $\left(g_{{\rm sp}}^{{\rm ac}} /v \right)^{2}$.  
This result is identical to our Eq.~(\ref{est_tau4}) and the
discussion is completely in line with the consideration given in
Sec.~\ref{Sec_ac}.

\subsubsection{Zone-boundary phonon regime and others}

We now consider the normal scattering processes from the optical-like,
zone-boundary phonons with ${\bf q}_{\parallel}\approx(\pi,\pi)$.  First,
we focus on temperatures $J\!\gg\!T\!\gg\!\omega_{0}\!\approx\!\Theta_{D}$.
Calculations are simplified by shifting the zero of the planar component of
the phonon momentum to the edge of the BZ, i.e. ${\bf q}_{ \parallel}
\rightarrow{ \bf q}_{ \parallel}+ ( \pi, \pi )$, and therefore $\mathbf{q }
\rightarrow \mathbf{q } +( \pi, \pi, 0)$. Since $T \! \ll \!J$, one can
still expand in (\ref{18}) with respect to small momenta $|\mathbf{k} +
\mathbf{q }_{\parallel}|$ and $|{\bf k}|$ near their respective ${\bf
k}$-points in the BZ, while for phonons one can set $\omega_{ \mathbf{
q}}^{ \ell} \! \sim \! \omega_{0}$ for all ${\bf q}$ involved in the
scattering. Using this expansion for vertices as discussed in Appendix
\ref{appA}, one can easily see that it leaves the structure of the
expression for the memory function in (\ref{22}) almost intact, with the
only difference that both ${\bf q}$ and ${\bf q}_{\parallel}$ are
large. Then, the power counting proceeds the same way as above with all the
variables related to magnons governed by the same smallness of the typical
momentum $k\sim T/v$, while for the phonon occupation number we now have
$n(\omega_{ \mathbf{ q}}^{ \ell}) \! \sim \! T/ \omega_{0}$ and the
summation over $q_{\perp}$ has no restrictions. Altogether, this yields
\begin{equation}
M_{\mu\nu}^{n}(i0^{+})\propto-i\,\delta_{\mu\nu}
\left(g_{{\rm sp}}^{{\rm zb}}\right)^{2}\,
\frac{T^{7}}{\omega_{0}v^{4}}\,,
\label{27}
\end{equation}
where $g_{{\rm sp}}^{{\rm zb}}\!=\!4S\lambda/\sqrt{2m\Theta_{D}}$ as given
in (\ref{Vp_opt1}). Similarly to the acoustic case, this implies the
following scattering rate
\begin{equation}
\frac{1}{\tau_{opt}^{n}}\propto\frac{T^{4}}{\omega_{0}v^{2}}\,,
\label{28}
\end{equation}
in a complete agreement with the relaxation rate obtained in
Sec. \ref{Sec_opt}, Eq. (\ref{est_tau2}).

Without repeating similar considerations for the anomalous scattering in
Fig.~\ref{fig1}(b), we simply note that the numerical evaluation to be
presented in the next subsection shows that for experimentally relevant
temperatures $T\gtrsim0.1\Theta_{D}$ its contribution is smaller by several
orders of magnitude as compared to normal scattering.

Finally, we note that in the unphysical regime of $T\!\gg\!J$, the
memory function scales as $M\sim T^{2}$ due to the three distribution
functions and the prefactor of $1/T$. Together with $\chi_{\mu\nu}\sim T$
for this regime, this yields a relaxation rate of $\tau^{-1}\propto T$.

\begin{figure}[tb]
\includegraphics[width=1.04\columnwidth]{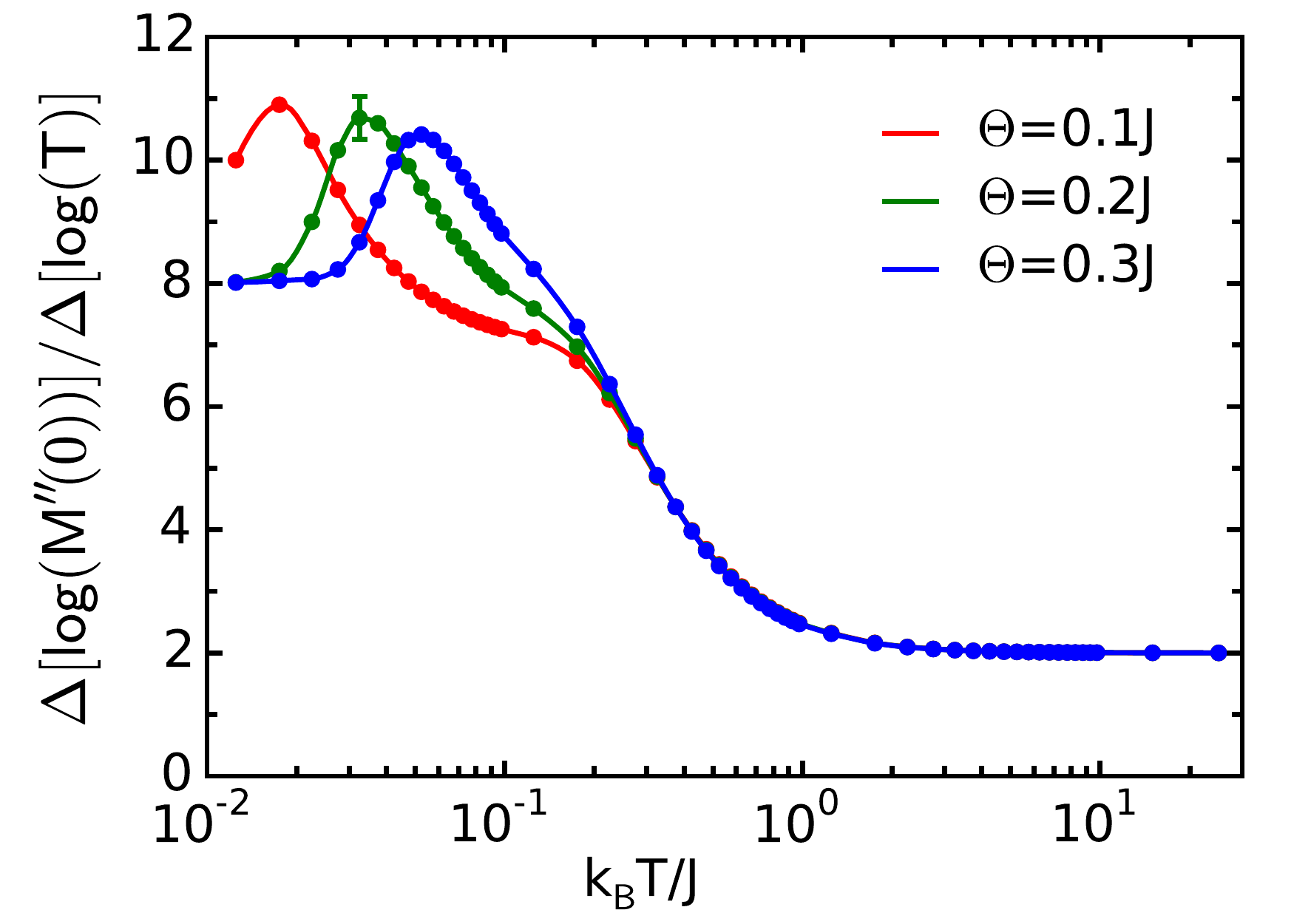} 

\protect\protect\caption{\label{fig2}(Color online)\
Difference quotient $n=\Delta\ln[M_{\mu\nu}(i0^{+})]/\Delta\ln T$ obtained
from numerical evaluation of $M_{\mu\nu}(i0^{+})$ versus $T$ (dots) for
various $\Theta_{D}$ (lines are guides to the eye).  Error bar is an
estimated maximum ``error'': $n-\partial\ln[M_{\mu\nu}(i0^{+})]/\partial\ln
T$.  For reference note that for $J=1550$K, $\Theta_{D}=0.1J$ corresponds
to a maximum phonon energy of approximately $270$K.}
\end{figure}

\subsection{Numerical analysis of the memory function}

Figure \ref{fig2} sheds light on our asymptotic analysis of relaxation
rates from an unbiased numerical point of view. The memory function has
been evaluated using Eqs.~(\ref{17}) and (\ref{20}) for three
representative Debye energies $\Theta_{D}$. The numerical integration has
been set up to satisfy the energy conserving delta-function exactly through
the numerical solution of $\varepsilon_{ \mathbf{k} + \mathbf{ q}} =
\varepsilon_{ \mathbf{k}} + \omega_{ \mathbf{ q}}^{ \ell}$ for each
integrand call. This leaves a 4D integration with a non-trivial
integrand. Evaluation of the latter has been performed for the temperatures
shown by the dots in Fig. \ref{fig2}. It shows a numerical approximation to
$\partial \ln [ M_{\mu\nu} (i0^{+}) ] / \partial \ln T$, the quantity which
allows to clearly identify the temperature ranges where the memory matrix
follows a power law, $M_{\mu\nu}( i0^{+}) \propto T^{n}$.  These ranges can
be seen as plateaus with the heights directly giving the exponent $n$.

We would like to stress several points. First, as predicted, a clear
$T^{8}$ regime, corresponding to $\tau^{-1}\propto T^{5}$, can be observed
for $T\ll\Theta_{D}$. However, comparing with the energy scales appropriate
for La$_{2}$CuO$_{4}$, such a regime is unlikely to be observed in the
current experimental studies of it, because the scattering in this
temperature range is dominated by the grain boundaries. \cite{Hess03}
Second, while this may be an artifact of our scheme of modeling the optical
phonon scattering by the phonon spectrum with a single phonon branch, for a
robust $T^{7}$ ($\tau^{-1}\propto T^{4}$) regime to occur, the Debye energy
needs to be low enough compared to $J$. This is evident from the set of
data in Fig.~\ref{fig2} with $\Theta_{D}=0.1J$ where clear indications of the
$T^{7}$ range can be observed. However, it is obvious from the results for
$\Theta_{D}=0.2J$ and $\Theta_{D}=0.3J$, that this regime rapidly merges with the onset of
thermalization of the high-energy magnons, where no well-defined exponent
can be observed. Third, the increase of the exponent between the $T^{8}$
and the $T^{7}$ regimes is consistent with an exponential increase of
$M_{\mu\nu}(i0^{+})$ with temperature. This is exactly the signature of the
Arrhenius (activated) behavior mentioned as regime (ii) in
Sec. \ref{Sec:asymptotes}. Finally, as expected for $T\gg J$ we get $M_{
\mu\nu} (i0^{+}) \propto T^{2}$.

\subsection{Thermal conductivity}

In this subsection we combine the memory function analysis of the spin-phonon 
coupling with the scattering mechanisms discussed in
Sec.~\ref{Sec_others} that are most significant for  magnon
thermal transport in large-$J$ antiferromagnets such as La$_{2}$CuO$_{4}$. 
Moreover, we will use parameters that may provide a reasonable
description of the experimental thermal conductivity data, e.g. of Ref.~\onlinecite{Hess03}. To
leading order, additional scattering mechanisms can be added according to the Matthiessen's rule, i.e.
summing their respective momentum-independent relaxations rates as in
Sec. \ref{sub:Relaxation-time-approximation}.

\subsubsection{Grain boundaries}

Grain boundary scattering is described in complete analogy with
Sec.~\ref{Sec_others_grain_boundary}, i.e. the results for thermal
conductivity trivially follow from Eq.~(\ref{14}) with the relaxation time
$\tau_{{\rm gb}}=l_{{\rm gb}}/v$, where $l_{{\rm gb}}$ is the typical grain
size and the magnon mean-free path, see (\ref{boundary}). As was
established in Ref.~\onlinecite{Hess03}, this scattering dominates thermal
conductivity of La$_{2}$CuO$_{4}$ at $T\!\alt\!200$K and one can show that
Eq.~(\ref{14}) with $l_{{\rm gb}}$ of order of a few hundred lattice
spacings and appropriate choice of $J$ and lattice parameters gives an
excellent quantitative description of the experimental $\kappa_{{\rm
gb}}(T)$ in this regime. For the remainder we choose $l_{{\rm gb}}=300$
lattice constants as in Sec.~\ref{RefCompar}

\begin{figure}[tb]
\includegraphics[width=1.04\columnwidth]{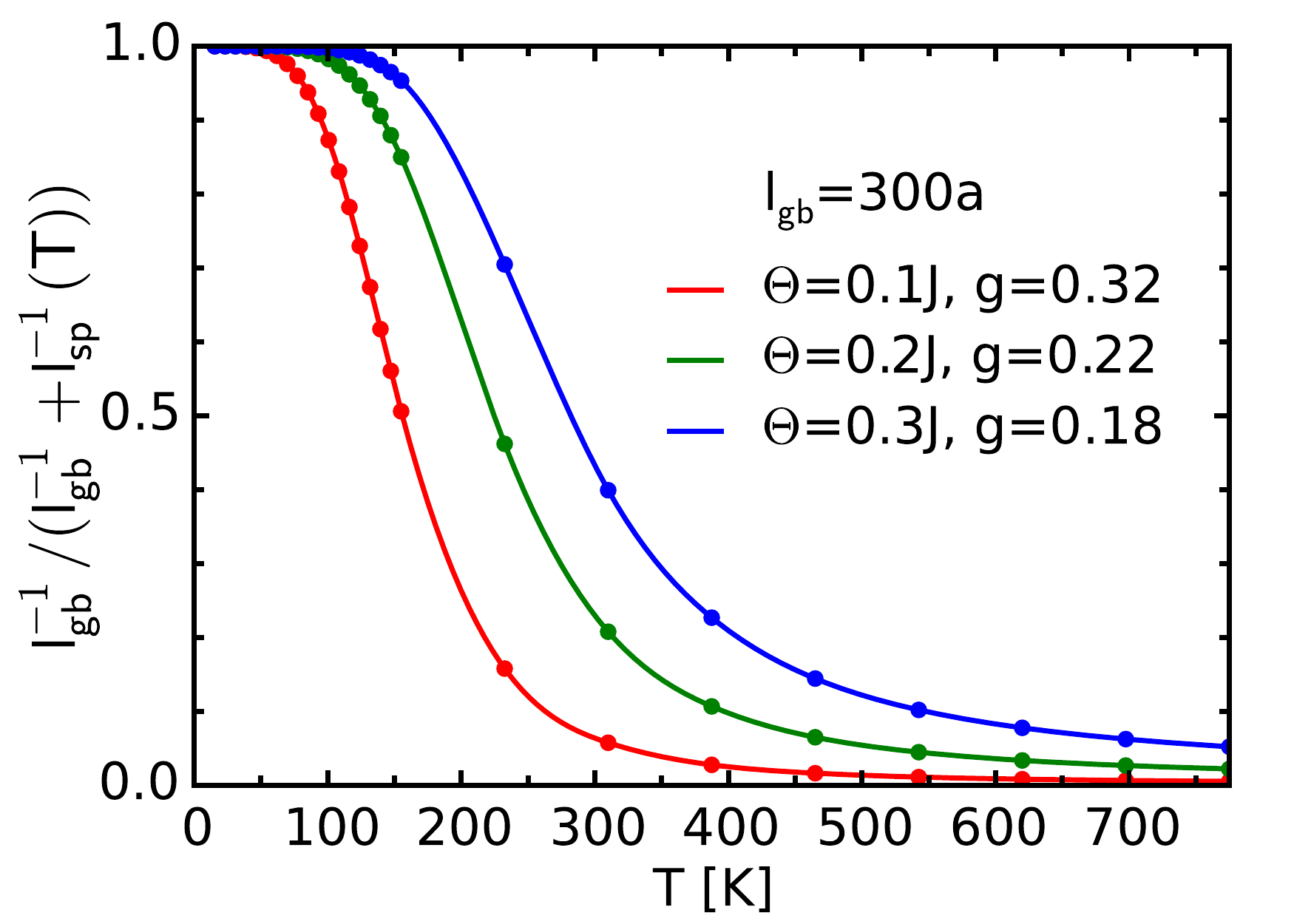} 
\protect\protect\caption{\label{fig4}(Color online) Renormalization factor of the 
inverse mean-free path due to grain boundaries 
from spin-phonon scattering for various $\Theta_{D}$ and $g$, $J=1550$K
(lines are guides to the eye).}
\end{figure}

\subsubsection{Grain boundaries and phonons}

We now would like to test if the magnon-phonon scattering can have
a significant impact on the magnon mean-free path and thermal conductivity.
For that, we calculate numerically the spin-phonon current relaxation
rate $(\mathbf{M}(i0^{+})/\boldsymbol{\chi})_{\mu\nu}=\delta_{\mu\nu}/\tau_{{\rm sp}}$
from (\ref{17}) as discussed above and neglecting anomalous scattering.
Then the spin-phonon transport mean-free path is defined as $l_{sp}=v\tau_{{\rm sp}}$
and can be combined with the grain-boundary mean-free path $l_{{\rm gb}}$
to see the effect of the former. To make such a comparison, we also
need to introduce a dimensionless spin-phonon coupling constant. In
line with the discussion of the asymptotic limits in Sec.~\ref{Sec:asymptotes}
above and in accord with the results of Appendix~\ref{appA}, such
a dimensionless coupling can be written as $g\!=\!4S\lambda/(J\sqrt{2m\Theta_{D}})$.
The range of its typical values is discussed in Sec.~\ref{Sec_ph_sum}
and Appendix~\ref{appD} and it is concluded that it may not exceed
unity.\cite{Bramwell90}
\begin{figure}[tb]
\includegraphics[width=1.04\columnwidth]{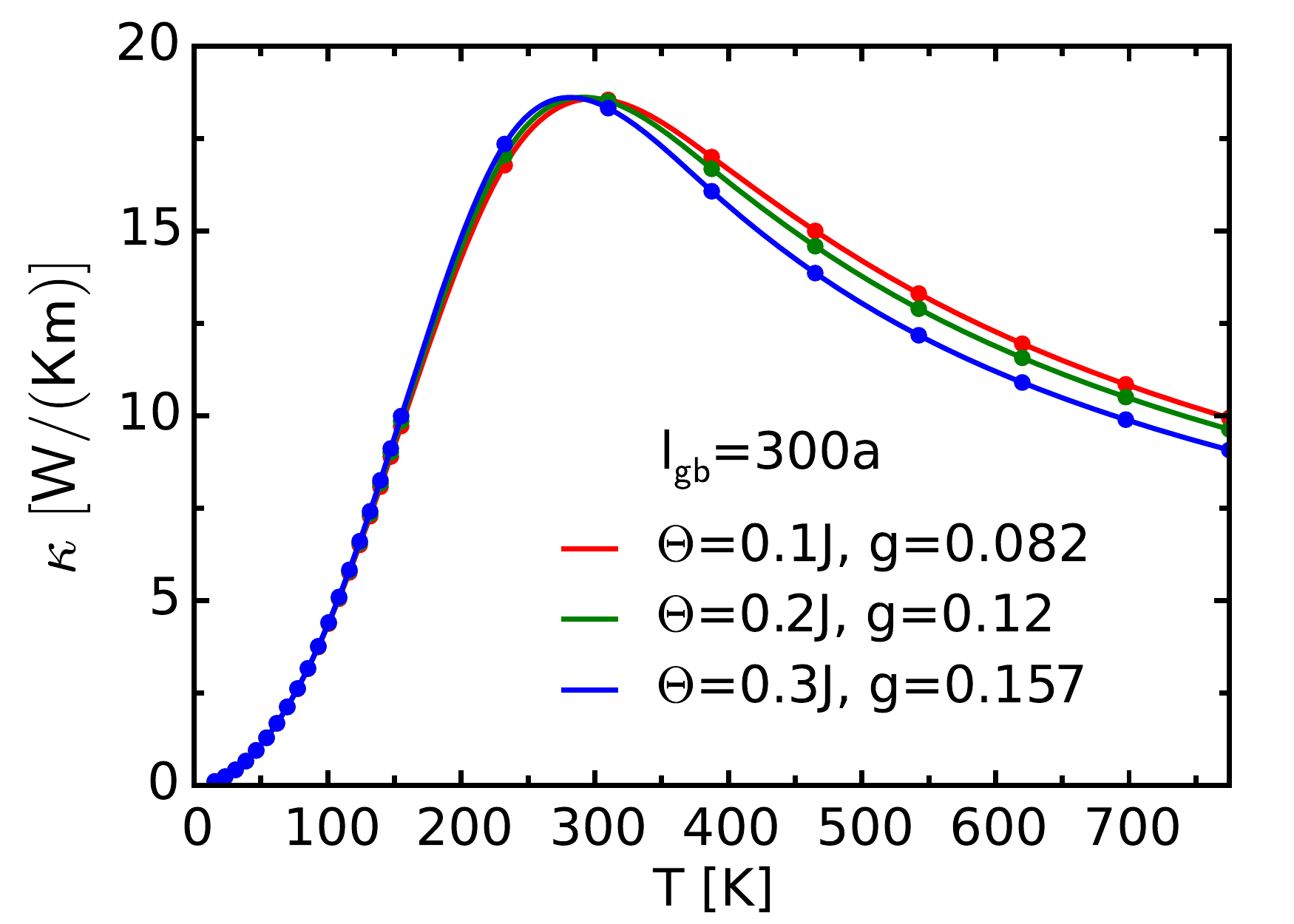} 
\protect\protect\caption{\label{fig5} (Color online) Magnon thermal conductivity
for the scattering by grain boundaries and phonons
for various experimentally reasonable values of $\Theta_{D}$ and $g$,
$J=1550$K (lines are guides to the eye).}
\end{figure}

Figure \ref{fig4} displays the ratio of the inverse magnon mean-free paths
with and without the spin-phonon scattering $l_{{\rm gb}}^{-1} / (l_{{\rm
gb}}^{-1}+l_{{\rm sp}}^{-1}(T))$ for a range of resonable spin-phonon
coupling constants and Debye temperatures. This Figure clearly demonstrates
that phonons can be expected to be the dominant scatterers for
$T\gtrsim200K$ in large-$J$ antiferromagnets with a modest spin-phonon
coupling.

In Figure \ref{fig5} the thermal conductivity is shown for the scenario
when only grain boundaries and phonons are involved in magnon scattering
and for several representative sets of $\Theta_{D}$ and $g$. The latter are
chosen to yield the same maximal value of $\kappa(T)$ at fixed $l_{{\rm
gb}}$, with $J$ and other parameters fixed to loosely match
La$_{2}$CuO$_{4}$. Hereafter all plots of the thermal conductivity 
display absolute values of $\kappa$. These follow from our calculations,
scaling the 2D conductivity to the 3D material \cite{Hess03}, and using 
$l_{gb} J k_B/(\hbar a c_z)\simeq 645$W/(Km), for $J\simeq k_B 1550$K and
$l_{gb}\simeq 300a$, with intra(inter)-plane lattice constants $a$($c_z$)
from \cite{Kastner98}. We emphasize, that the magnitude of $\kappa$ shown in
Fig. \ref{fig5} is  within the typical range for La$_{2}$CuO$_{4}$.
\cite{Hess03,others}  The Figure shows that the spin-phonon coupling
constants necessary to effectively suppress the conductivity at
$T\agt\Theta_{D}$ are well within the acceptable values. Moreover, the
Figure demonstrates that a rather natural temperature range for the maximum
conductivity to occur in La$_{2}$CuO$_{4}$ and related materials is of the
order of $\Theta_{D}$. Comparing with Fig.~\ref{fig2}, one can see that the
decrease of $\kappa$ vs $T$ occurs in the temperature regime where no
well-defined power law exists for $\tau_{{\rm sp}}^{-1}(T)$, independently of
the choice of the Debye energy.

Fig.~\ref{fig5} should also be compared with the upper curve in
Fig.~\ref{Fig_kappa}, which is obtained from the effective
$1/\tau$-approximation within the Boltzmann formalism. Taking into
account the differences between the $1/\tau$-approximation in Boltzmann approach
and the memory function calculations, 
different types of modeling of the phonon bath, and  keeping in mind
differences in Debye energies and spin-phonon coupling
constants, the close similarity of the overall shape and other features
of the $\kappa(T)$ curves from the Boltzmann and the memory function approaches
in Figs.~ \ref{Fig_kappa} and \ref{fig5} are remarkable.

\begin{figure}[tb]
\includegraphics[width=1.04\columnwidth]{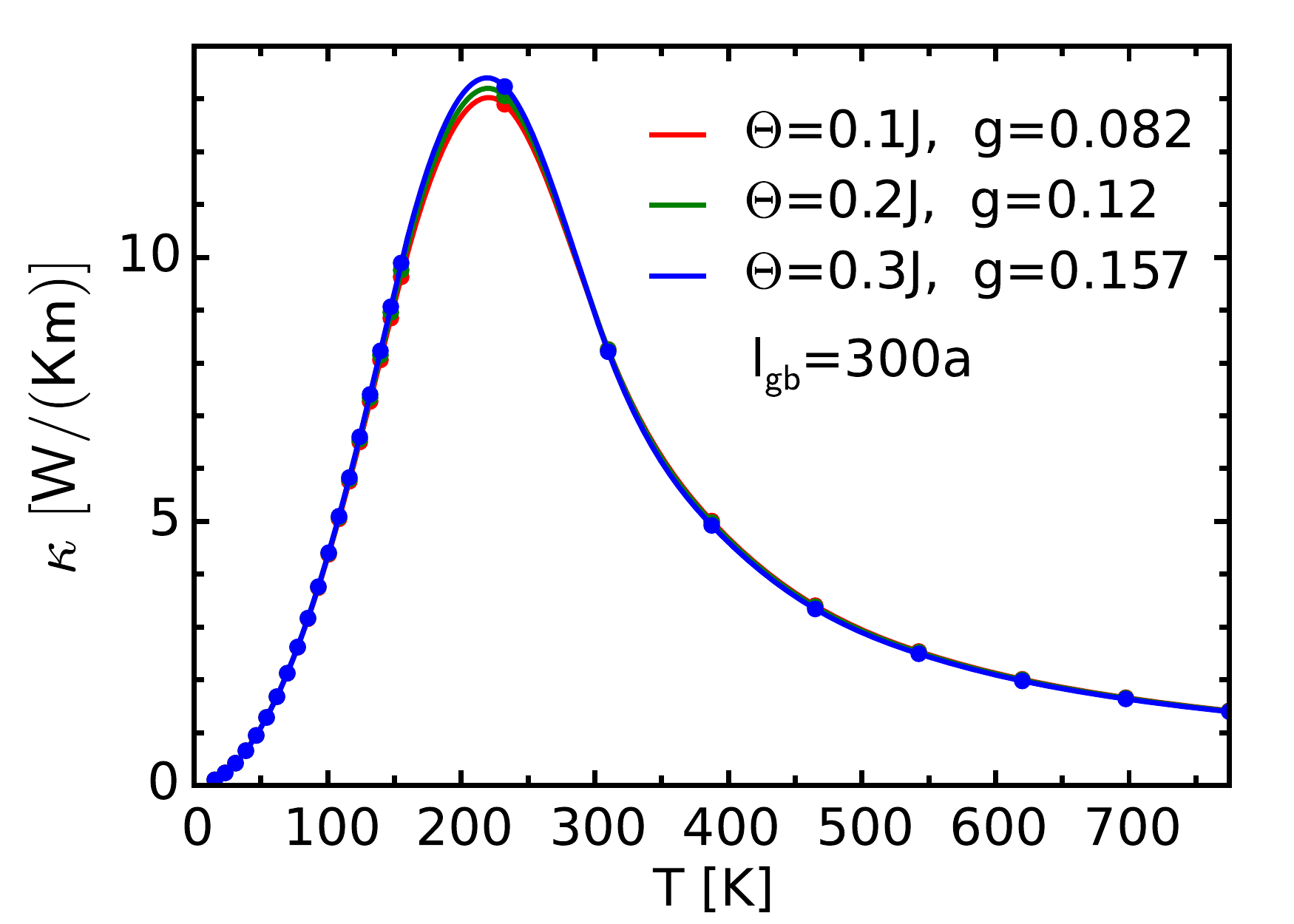} 
\protect\protect\caption{\label{fig7} (Color online) Magnon thermal conductivity
for the  scattering off the effective grain boundary due to temperature-dependent correlation length 
and phonons for various experimentally reasonable values of $\Theta_{D}$ and $g$, $J=1550$K (lines are guides to the eye).}
\end{figure}

\subsubsection{Grain boundaries, phonons, and finite correlation length}

Next we take into account scattering due to the finite spin-spin
correlation length in the paramagnetic phase above the ordering N\'{e}el
temperature, analogous to the discussion in Sec.~\ref{Sec_others}. Similar
to the Boltzmann approach, a formally proper treatment of its effects is
beyond the present study. A qualitative description, however, can be obtained
readily. First, a finite correlation length implies a mass gap in the
magnon dispersion, discussed in Sec.~\ref{Sec_others}, see
Eq.~(\ref{Ek_xi}).\cite{Takahashi} Second, since the notion of magnetic
order is meaningful only within patches of linear dimension $\sim\xi(T)$,
Eq.~(\ref{xi}), the system consists of effective ``grains'' with a
\emph{temperature-dependent}  size $l_{gb}(T) \approx \mathrm{min}
(l_{{\rm gb}}, \xi(T))$, the  sentiment expressed earlier in
Sec.~\ref{boltzmann}. Once again, Matthiessen's rule is used to yield a
smooth crossover  between the two
grain-boundary  regimes as a function of temperature and to include the spin-phonon
scattering with the inverse mean-free path given by $l^{-1}(T) =
l^{-1}_{gb} + \xi^{-1}(T) + l^{-1}_{sp}$.

First, recalculating $\kappa$ within the relaxation time approximation,
Eq. (\ref{14}), with a constant scattering time $\tau$ and the mass gap  due to correlation-length 
in the magnon dispersion, Eq. (\ref{Ek_xi}), affects $\kappa$ only weakly and only at higher
temperatures, in a broad agreement with the conclusions reached in
Sec.~\ref{Sec_kappa}. For brevity, we will not study the impact of the mass
gap on the memory function. In turn, the main effect of the correlation length on
the thermal conductivity is from the limiting of the mean-free path by
$\xi(T)$. For the parameters relevant to the cuprates, the correlation length
gets short enough to likely dominate dominate the thermal conductivity for
$T\gtrsim250$K.

\begin{figure}[tb]
\includegraphics[width=1.04\columnwidth]{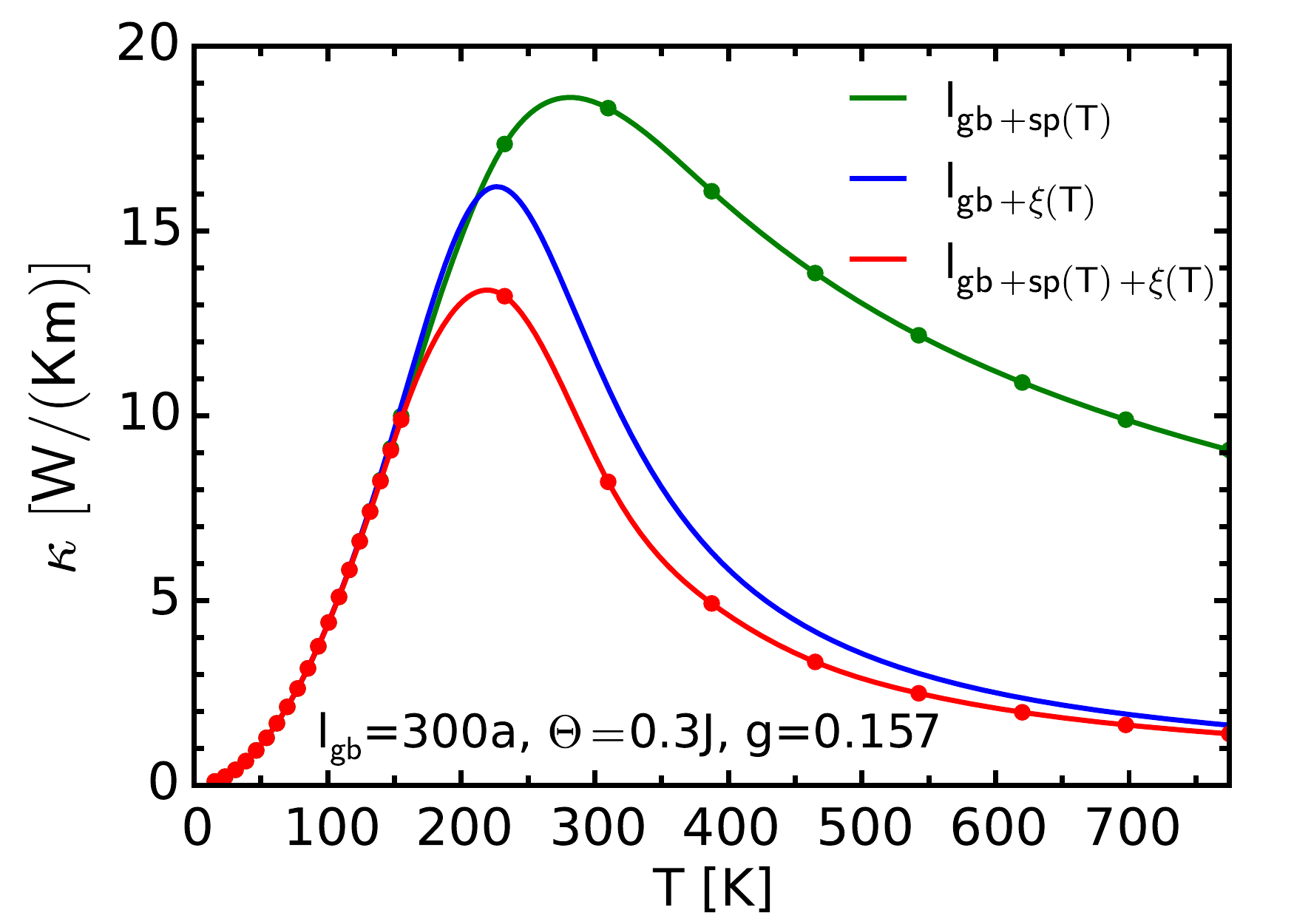} 
\protect\protect\caption{\label{fig8} (Color online) Contrasting impact of
different scattering mechanisms on the thermal conductivity. Phonons and crystalline grain boundary (upper curve), 
correlation length and crystalline grain boundary (middle curve), phonons, correlation length,
and crystalline grain boundary (lower curve).}
\end{figure}

Our Figures \ref{fig7} and \ref{fig8} display the combined effect of
spin-phonon and effective boundary scattering, discarding the magnon mass
gap. Fig. \ref{fig7} shows that at high temperatures,  in contrast to
the rather slow decrease of thermal conductivity in Fig.~\ref{fig5}, $\kappa(T)$
is dominated by the exponential decrease of the effective grain size $\xi(T)$. This figure also shows  
that small differences in the results  Fig.~\ref{fig5}
due to differences in coupling constants and Debye energies  are completely suppressed by inclusion of $\xi(T)$. 
Fig. \ref{fig8} contrasts the various scattering mechanisms against each
other. This figure clearly demonstrates that while the spin-spin
correlation length seems to provide the major limit on $\kappa(T)$ at
higher temperatures, magnon-phonon scattering still contributes
significantly to the suppression of the magnon heat current up to
high temperatures.

Once more we emphasize the close similarity of the combined results from
the memory function method in Fig. \ref{fig8} with those from the Boltzmann
theory in Fig.~\ref{Fig_kappa2}. Given that these results have been
obtained independently from two  distinct theoretical approaches, this agreement
strongly corroborates our main conclusions.

\section{Summary}

To summarize, we have considered relaxation mechanisms of 2D magnons in
large-$J$ antiferromagnets such as La$_2$CuO$_4$ with the goal of providing
a basis for quantitative calculations of thermal transport by spin degrees
of freedom in this class of materials. We conclude that the magnon thermal
conductivity in these and related systems is limited by three key
scattering mechanisms: grain boundaries, 2D correlation length, and
magnon-phonon scattering, the latter being effective at intermediate and
high temperatures, $T\!\agt\!\Theta_D$. Impurity-like lattice-induced
disorder and magnon Umklapp scattering have been found to be less important.

The bulk of our consideration has been devoted to the scattering of 2D
magnons on 3D phonons, with acoustic and optical phonon branches examined
in detail.  Within the Boltzmann approach and $1/\tau$ approximation we
have advocated the use of a simplified ``effective'' phonon DoS approach,
which allows for straightforward yet fairly realistic calculations yielding
an effective expression for the magnon relaxation rate on optical and
acoustic phonons that contains minimal number of parameters.  We have also
employed the memory function approach, in which we have retained the full
microscopic expressions for the magnon and phonon spectra, with the only
approximation being the coupling of spins to only a single
dispersive phonon branch.

Within both approaches, we have closely analyzed the power-law asymptotic
regimes of the magnon relaxation rates involving acoustic and optical, or
zone-boundary, phonons.  Not only have such investigations proven very
instructive in the analysis of which part of the magnon population is
affected most strongly by phonons, but also have demonstrated a remarkable
accord between the two very different theoretical approaches to the
transport problem.

In both Boltzmann and memory functions approaches, we have demonstrated
that having a modest spin-phonon coupling, well within a reasonable range
of parameters, phonons can have a significant impact on the magnon
mean-free path and thermal conductivity.  Taking into account other
scattering mechanisms within the thermal conductivity calculations, we have
concluded that magnon thermal conductivity in large-$J$ antiferromagnets
should be largely controlled by the ``boundary-like'' scatterings, coming
from either the real grain boundaries or the 2D correlation length, with a
substantial correction from the magnon-phonon scattering, affecting
$\kappa(T)$ at intermediate and high temperatures, $T\!\agt\!\Theta_D$.  We
have also demonstrated that the results of both approaches on thermal
conductivity are remarkably similar.

\begin{acknowledgments}

We acknowledge numerous enlightening discussions with Christian Hess.  Work
of A.~L.~C. was supported by the U.S. Department of Energy, Office of
Science, Basic Energy Sciences under Award \# DE-FG02-04ER46174.
A.~L.~C. would like to thank Aspen Center for Physics, where part of this
work was done, for hospitality. The work at Aspen was supported in part by
NSF Grant No. PHYS-1066293.  Work of W.~B. has been supported in part by the
European Commission through the LOTHERM project \# PITN-GA-2009-238475, by
the Deutsche Forschungsgemeinschaft through SFB 1143 and FOR 912 Grant No.\
BR 1084/6-2, as well as by the Lower Saxony PhD program SCNS.  W.~B. also
acknowledges kind hospitality of the Platform for Superconductivity and
Magnetism, Dresden.
\end{acknowledgments}

\appendix

\section{Spin model}
\label{app0}

For the spin system we consider the two-dimensional,  nearest-neighbor, 
spin-1/2 Heisenberg antiferromagnet on a square-lattice
\begin{equation}
{\cal H}_{\rm s}=J\sum_{\langle ij\rangle} {\bf S}_i \cdot{\bf S}_j\,,\label{1}
\end{equation}
where summation is over the nearest-neighbor bonds  and
$J$ is the exchange coupling constant.  We treat this model using standard linear
spin-wave theory. 
For the bipartite lattice we 
transform from the laboratory to the rotated frame for spins and  
bosonize spin operators according to 
\begin{eqnarray}
&&S^{z_0}_i=e^{i{\bf Q}{\bf r}_i} S^z_i, \ \ S^{x_0}_i=e^{i{\bf Q}{\bf r}_i} S^x_i, \ \ S^{y_0}_i=S^y_i, \ \ \ \ 
\label{HP}\\
&& S^z_i=S-b_i^\dag b_i, \ \ S_i^+=\left(S_i^-\right)^\dag\approx b_i\sqrt{2S}
\, , \nonumber
\end{eqnarray}
where ${\bf Q}=(\pi,\pi)$ is the antiferromagnetic ordering vector.
Then, within the  $1/S$ approximation, one needs to retain only quadratic terms
in the  Hamiltonian (\ref{1}) 
\begin{eqnarray}
{\bf S}_i \cdot{\bf S}_{i+\delta}\Rightarrow S\Big\{b_i^\dag b_i  + b_{i+\delta}^\dag b_{i+\delta}-
b_{i+\delta}^\dag b_{i}^\dag-b_{i+\delta} b_{i} \Big\} .\ \ 
\label{SS}
\end{eqnarray}
After the Fourier transform, the subsequent treatment of (\ref{1}) involves Bogolyubov transformation 
of magnons given by 
\begin{eqnarray}
b_{\bf k}\!=\!u_{\bf k} \beta_{\bf k}\!+\!v_{\bf k}\beta^\dag_{\bf -k}\, ,
\label{Bogo0}
\end{eqnarray}
with
\begin{eqnarray}
u_{\bf k}=\sqrt{\frac{1+\nu_{\bf k}}{2\nu_{\bf k}}}, \ \ 
v_{\bf k}=sign\left(\gamma_{\bf k}\right)\sqrt{\frac{1-\nu_{\bf k}}{2\nu_{\bf k}}}, 
\label{Bogo}
\end{eqnarray}
where $\nu_{\bf k}\!=\!\sqrt{1-\gamma_{\bf k}^2}$ is related to the 
  magnon energy via $\varepsilon_{\bf k}\!=\!4JS\nu_{\bf k}$ 
 and $\gamma_{\bf k}\!=\!\left(\cos k_x\!+\!\cos k_y\right)/2$.
Finally, the spin-wave Hamiltonian is given by
\begin{equation}
{\cal H}_{\rm s}^{(2)}=\sum_{\mathbf{k}}\varepsilon_{\mathbf{k}}
\beta_{\mathbf{k}}^{\dagger}\beta_{\mathbf{k}}^{\vphantom{\dagger}}
\,,\label{3}
\end{equation}
At long wavelength
$\varepsilon_{\mathbf{k}}\approx vk$ with the spin-wave velocity
$v=\sqrt{2}J$.

\section{Microscopic derivation of spin-phonon Hamiltonians: vertices, polarizations, etc.}
\label{appA}

\subsubsection{Spin-phonon Hamiltonian: Bravias lattice}

For an isotropic, nearest-neighbor  interaction 
of spins and considering the lattice of only magnetic ions,
the most natural source of the spin-phonon coupling is from the modifications of the superexchange due to local 
stretching or compression of the bond length\cite{Dixon}
\begin{eqnarray}
&&{\cal H}_{\rm s-ph}=\sum_{\langle ij\rangle}\delta J\left({\bf r}_j-{\bf r}_i\right) {\bf S}_i \cdot{\bf S}_j
\label{Hsph0}\\
&&\phantom{{\cal H}_{\rm s-ph}}\approx\frac{\lambda }{2}
\sum_{i,\delta}  
\left(\mbox{\boldmath$\delta$}\cdot\Delta{\bf U}_{i,\delta}\right){\bf S}_i \cdot{\bf S}_{i+\delta}
\, , \nonumber
\end{eqnarray}
where $\lambda\!=\!a\partial J/\partial r$, with $a$ being the lattice constant, 
$\delta$ runs over the nearest neighbors,  {\boldmath$\delta$} are the 
corresponding unit vectors of the lattice, $\Delta{\bf U}_{i,\delta}\!\equiv\!\left({\bf U}_{i+\delta}\!-\!{\bf U}_i\right)$, 
and ${\bf U}_i$ is the displacement vector of the $i$th ion. 
While the Hamiltonian in (\ref{Hsph0}) is rather general, one can see it as describing a single 2D plane of spins
on a square lattice, representative of a CuO$_2$ plane of La$_2$CuO$_4$. Below, we will extend this picture to a 
tetragonal array of magnetically decoupled planes to take into account the 3D nature of phonons.
Note that while the displacement of ions are described by 3D vectors,  in (\ref{Hsph0}) only their projections on the in-plane
bonds ({\boldmath$\delta$}) matter. 

After bosonizing spin operators according to (\ref{SS}), 
using  the symmetry of the square lattice,  
Fourier transform, and some algebra, we obtain
\begin{eqnarray}
&&{\cal H}_{\rm s-ph}=4i\lambda S\sum_{\bf q, k, k'} \sum_{\mbox{\boldmath$\delta$}}
\left(\mbox{\boldmath$\delta$}\cdot{\bf U}_{\bf q}\right)
\sin\left(\frac{{\bf k}-{\bf k'}}{2}{\mbox{\boldmath$\delta$}}\right)
\nonumber \\
&& \phantom{{\cal H}=4i\lambda S}
\bigg\{\cos\left(\frac{{\bf k}-{\bf k'}}{2}{\mbox{\boldmath$\delta$}}\right) b_{\bf k'}^\dag b_{\bf k} 
\label{Hsph1}\\
&& \phantom{{\cal H}=4i\lambda S}
-\frac12 \cos\left(\frac{{\bf k}+{\bf k'}}{2}{\mbox{\boldmath$\delta$}}\right)
\left( b_{\bf k'}^\dag b_{-\bf k}^\dag+b_{-\bf k'} b_{\bf k} \right)\bigg\}
\, , \nonumber
\end{eqnarray}
where ${\bf q}\!=\!({\bf k}-{\bf k'}, q_\perp)$ and the
 summation over the directions of the bond now takes the values of only two 
unit vectors, $\mbox{\boldmath$\delta$}\!=\! {\bf \hat{x}}$ and ${\bf \hat{y}}$. In derivation of (\ref{Hsph1})
we have performed summation over the sites of the tetragonal lattice of magnetically decoupled planes. It is natural that the
in-plane component of the phonon momentum (in the index of ${\bf U}_{\bf q}$) is tied to the magnon momenta via momentum
conservation, ${\bf q}_\parallel\!=\!{\bf k}-{\bf k'}$, while the component of the phonon momentum 
perpendicular to the plane, ${\bf q}_\perp$, is not conserved and is an independent variable in the sum 
in (\ref{Hsph1}). This is because the 2D magnons have zero coupling between the planes 
(infinite mass in that direction) and thus provide no restriction on  ${\bf q}_\perp$. 

The subsequent treatment of (\ref{Hsph1}) involves Bogolyubov transformation of magnons as in (\ref{Bogo0}) and 
 rewriting the displacement operators in terms of phonons. 
For the Bravias lattice, the ${\bf q}$'s  Fourier component of the lattice displacement is given by
\begin{eqnarray}
{\bf U}_{\bf q}=\sum_\ell \frac{\mbox{\boldmath$\xi$}^\ell_{\bf q}}{\sqrt{2m\omega^\ell_{\bf q}}}
\left(a^\dag_{{\bf q}\ell}+a_{-{\bf q}\ell}\right),
\label{Uq}
\end{eqnarray}
where $\ell=1,2,3$ numerates one longitudinal and two transverse phonon polarizations, 
{\boldmath$\xi$}$^\ell_{\bf q}$ are the polarization unit vectors, $m$ is the mass of the unit cell,
and $\omega^\ell_{\bf q}$ and $a_{{\bf q}\ell}$ are the energies and operators of the corresponding 
phonon branches, respectively.

Applying (\ref{Uq}) and (\ref{Bogo}) to (\ref{Hsph1}) we obtain
\begin{eqnarray}
&&{\cal H}_{\rm s-ph}=\sum_{{\bf k, k'}, q_\perp} \sum_{\ell}
\Big\{V^{\ell}_{\bf k, k', q} \beta_{\bf k'}^\dag \beta_{\bf k} 
\label{Hsph2}\\
&& \phantom{{\cal H}=}
+\frac12\, V^{{\rm od},\ell}_{\bf k, k', q}\left( \beta_{\bf k'}^\dag \beta_{-\bf k}^\dag+\mbox{H.c.} \right)\Big\}
\left(a^\dag_{{\bf q}\ell}+a_{-{\bf q}\ell}\right), 
\nonumber
\end{eqnarray}
where ``normal'' and ``anomalous'' magnon-phonon vertices are
\begin{eqnarray}
&&V^{\ell}_{\bf k, k', q} = \frac{4S\lambda}{\sqrt{2m\omega^\ell_{\bf q}}}
\sum_{\alpha=x,y} \xi^{\ell,\alpha}_{\bf q}  \sin\frac{q_{-}^\alpha}{2}
\label{Vmagph}\\
&& \phantom{V^{\ell}_{\bf k, k', q}}
\times\Big\{\cos\frac{q_{-}^\alpha}{2} \left(uu^\prime + vv^\prime\right)-
\cos\frac{q_{+}^\alpha}{2}\left(uv^\prime + vu^\prime\right)\Big\},
\nonumber \\
&& 
V^{{\rm od},\ell}_{\bf k, k', q}= \frac{4S\lambda}{\sqrt{2m\omega^\ell_{\bf q}}}
\sum_{\alpha=x,y} \xi^{\ell,\alpha}_{\bf q}  \sin\frac{q_{-}^\alpha}{2}
\label{Vmagph1}\\
&& \phantom{V^{{\rm od},\ell}_{\bf k, k', q}}
\times\Big\{\cos\frac{q_{-}^\alpha}{2} \left(uv^\prime + vu^\prime\right)-
\cos\frac{q_{+}^\alpha}{2}\left(uu^\prime + vv^\prime\right)\Big\},
\nonumber
\end{eqnarray}
where we have absorbed $i$ in the definition of phonon operators,
$\xi^\alpha$ and $q_{\pm}^\alpha$ are the $x$ and $y$ projections of the corresponding 
vectors, and we have introduced shorthand notations ${\bf q}_{\pm}\!=\!{\bf k}\!\pm\!{\bf k'}$,
$u^{(\prime)}\!=\!u_{\bf k^{(\prime)}}$, and $v^{(\prime)}\!=\!v_{\bf k^{(\prime)}}$.

\subsubsection{Asymptotics and polarizations}

Of interest, of course, is the asymptotic behavior of these vertices in the vicinity of 
${\bf k}, {\bf k'}\!\rightarrow\! \mbox{\boldmath$\Gamma$}, {\bf Q}$ as the momenta of magnons are confined to these
regions at not too high temperatures.

Expanding (\ref{Vmagph}) for magnon momenta 
${\bf k}, {\bf k'}\!\rightarrow\! \mbox{\boldmath$\Gamma$}$ (or ${\bf Q}$), which also corresponds
to ${\bf q}_\parallel \!\rightarrow\! 0$, gives $\alpha=x,y$ components of the vertex
\begin{eqnarray}
V^{\ell,\alpha}_{\bf k, k', q} \approx \frac{S\lambda\,\xi^{\ell,\alpha}_{\bf q}}{\sqrt{4m\omega^\ell_{\bf q}}}
\ q^\alpha_\parallel \sqrt{|{\bf k}||{\bf k'}|}\,\left(1+\frac{2k^\alpha k'^\alpha}{|{\bf k}||{\bf k'}|}\right),
\label{Vmagph2}
\end{eqnarray}
and expansion of (\ref{Vmagph1}) yields $V^{{\rm od},\ell,\alpha}_{\bf k, k', q}=-V^{\ell,\alpha}_{\bf k, k', q}$.

In general,  three polarization vectors {\boldmath$\xi$}$^\ell_{\bf q}$ 
in (\ref{Uq}) should be obtained as solutions of the dynamical matrix 
equation for the lattice vibrations of the tetragonal Bravias lattice for each ${\bf q}$-vector. 
However, in the long-wavelength limit ${\bf q}\!\rightarrow\! 0$, these solutions can be classified 
as in the continuum as one longitudinal and two transverse modes. For the former, 
{\boldmath$\xi$}$^{(3)}_{\bf q}\!\approx\!{\bf q}/|{\bf q}|$ is along the momentum of the phonon, while the 
latter can be chosen freely as an orthogonal pair in the plane perpendicular to ${\bf q}$. For instance, a convenient choice
is {\boldmath$\xi$}$^{(1)}_{\bf q}\!\approx\!{\bf y}\times{\bf q}/|{\bf y}\times{\bf q}|$, which ensures that 
it is orthogonal to the $y-q$ plane. Since in (\ref{Vmagph})--(\ref{Vmagph2})
we are interested only in $x$ and $y$ projections of the polarization vectors,  this choice leaves nonzero
only $\xi^{(1),x}_{\bf q}$ component. 

However, the situation is even simpler because, according to the discussion in Sec.~\ref{prelim}, 
in a typical scattering process component of the phonon momentum perpendicular to the plane 
is much larger than the in-plane component,
$|{\bf q}_\perp|\!\gg\!|{\bf q}_\parallel|$. This means that the in-plane projections of the 
longitudinal phonon polarization vector are negligible, 
$\xi^{(3),x(y)}_{\bf q}\!\approx\!q_\parallel^{x(y)}/q_\perp\!\ll\!1$. 
In turn, this means that the polarizations of the two transverse modes lie almost entirely in the $x-y$ plane
and the corresponding vectors can simply be chosen along the $x$ and $y$ axes with 
$\xi^{(1),x}_{\bf q}\!\approx\!\xi^{(2),y}_{\bf q}\!\approx\! 1$ and 
$\xi^{(1),y}_{\bf q}\!\approx\!\xi^{(2),x}_{\bf q}\!\approx\! 0$. Therefore, it is the transverse phonons with 
the momentum largely orthogonal to the antiferromagnetic planes that couple most strongly to the 
spin excitations. As a result, summation over $\ell$ in (\ref{Hsph2}) should concern only them. 
We note that the same arguments have been used previously for the spin-phonon coupling in
1D spin chains, see Ref.~\onlinecite{RC}.

Thus, using $\omega^{(1)}_{\bf q}\!\approx\!\omega^{(2)}_{\bf q}\!\approx\! c|{\bf q}|$ 
for the transverse branches,  the magnon-phonon coupling with the acoustic phonons is reduced to just 
two non-zero terms
\begin{eqnarray}
&&V^{(1),x}_{\bf k, k', q} \approx g_{\rm sp}^{\rm ac} \, \frac{q^x_\parallel}{\sqrt{|{\bf q}|}}
\,  \sqrt{|{\bf k}||{\bf k'}|}\,\left(1+\frac{2k^x k'^x}{|{\bf k}||{\bf k'}|}\right), \ \ \ 
\label{Vmagph3}\\
&&V^{(2),y}_{\bf k, k', q} \approx g_{\rm sp}^{\rm ac} \, \frac{q^y_\parallel}{\sqrt{|{\bf q}|}}
\,  \sqrt{|{\bf k}||{\bf k'}|}\,\left(1+\frac{2k^y k'^y}{|{\bf k}||{\bf k'}|}\right),\nonumber
\end{eqnarray}
where $g_{\rm sp}^{\rm ac}\!=\!S\lambda/2\sqrt{mc}$ and the results are the same for 
the anomalous coupling $V^{{\rm od},\ell,\alpha}_{\bf k, k', q}$. 

Since the vertices in (\ref{Vmagph3}) couple to different but degenerate phonon branches, they will 
contribute independently to the scattering probability, so that $|V_{\rm tot}|^2\!=\! |V^{(1),x}|^2\!+\!|V^{(2),y}|^2$. 
Then it is natural to introduce 
an effective coupling to just one phonon branch that would lead to the same scattering rate  
\begin{eqnarray}
V^{\rm eff}_{\bf k, k', q} \approx g_{\rm sp}^{\rm ac} \, \frac{|{\bf q}_\parallel|}{\sqrt{|{\bf q}|}}
\,  \sqrt{|{\bf k}||{\bf k'}|}\, ,
\label{Vmagph4}
\end{eqnarray}
in which we simply ignored the angular dependence in the brackets in (\ref{Vmagph3}) [second terms]. This 
is equivalent only to a quantitative (order of 1) change in the effective coupling constant.
Needless to say, the result in (\ref{Vmagph4}) is identical to the coupling proposed in (\ref{Vac}) and
used throughout the paper.

With that the effective Hamiltonian for magnon-phonon coupling becomes 
\begin{eqnarray}
&&{\cal H}^{\rm eff}_{\rm s-ph}=\sum_{{\bf k, k'}, q_\perp} 
V^{\rm eff}_{\bf k, k', q}\Big\{\beta_{\bf k'}^\dag \beta_{\bf k} 
+\frac12\left( \beta_{\bf k'}^\dag \beta_{-\bf k}^\dag+\mbox{H.c.} \right)\Big\}
\nonumber\\
&& \phantom{{\cal H}_{\rm eff}=\sum_{{\bf k, k'}, q_\perp} V^{\rm eff}_{\bf k, k', q}}
\times\left(a^\dag_{{\bf q}\ell}+a_{-{\bf q}\ell}\right), 
\label{Heff1}
\end{eqnarray}

\subsubsection{Zone-boundary phonon}

A yet another asymptotic consideration is relevant for the magnon-phonon
coupling on a Bravias lattice. It concerns scattering of a magnon between two branches, from the vicinity of 
{\boldmath$\Gamma$} point to  the gapless branch at the antiferromagnetic ordering vector ${\bf Q}$ or vice versa.
Therefore, of interest are the limits ${\bf k}\!\rightarrow\! \mbox{\boldmath$\Gamma$} $ 
and  ${\bf k'}\!\rightarrow\! {\bf Q}$ for vertices in (\ref{Vmagph}) and (\ref{Vmagph1}). 
Importantly, the phonon involved in such a process
has a large momentum with the in-plane component ${\bf q}_\parallel\!\approx\!-{\bf Q}$, which corresponds to 
a zone-boundary excitation with the energy of order $\Theta_{\rm D}$ and  is similar to the optical
modes considered next.

Expansion of the ``magnon part'' [content of the curly brackets] in (\ref{Vmagph}) and (\ref{Vmagph1})
gives 
\begin{eqnarray}
&&V^{\ell}_{\bf k, k', q} = \frac{4S\lambda}{\sqrt{2m\omega^\ell_{\bf q}}}
\sum_{\alpha} \xi^{\ell,\alpha}_{\bf q}\bigg\{\frac{k'^\alpha|{\bf k}|-k^\alpha|{\bf k'}|}{2\sqrt{|{\bf k}||{\bf k'}|}}\bigg\},
\ \ \ \ \ \ \ 
\label{Vp_opt}\\
&& 
V^{{\rm od},\ell}_{\bf k, k', q}= \frac{4S\lambda}{\sqrt{2m\omega^\ell_{\bf q}}}
\sum_{\alpha} \xi^{\ell,\alpha}_{\bf q}\bigg\{\frac{k^\alpha|{\bf k'}|+k'^\alpha|{\bf k}|}{2\sqrt{|{\bf k}||{\bf k'}|}}\bigg\},
\nonumber
\end{eqnarray}
where we have shifted ${\bf k'}\!\rightarrow\!{\bf Q}\!+\!{\bf k'}$  and 
also used that $\sin(q^\alpha_\parallel/2)\!\approx\! -1$ for ${\bf q}_\parallel\!\approx\!-{\bf Q}$.
Focusing now on the ``normal'' vertex and following the same choice of polarization vectors as above 
leads to two non-zero vertices
\begin{eqnarray}
&&V^{(1),x}_{\bf k, k', q} \approx g_{\rm sp}^{\rm zb} \, 
\frac{k'^x|{\bf k}|-k^x|{\bf k'}|}{2\sqrt{|{\bf k}||{\bf k'}|}}\, ,
\label{Vp_opt1}\\
&&V^{(2),y}_{\bf k, k', q} \approx g_{\rm sp}^{\rm zb} \, 
\frac{k'^y|{\bf k}|-k^y|{\bf k'}|}{2\sqrt{|{\bf k}||{\bf k'}|}}\, ,
\nonumber
\end{eqnarray}
with the coupling constant to the zone-boundary phonon $g_{\rm sp}^{\rm zb}\!=\!4S\lambda/\sqrt{2m\Theta_{\rm D}}$.
Combining the two couplings to different phonon branches into one effective vertex gives
\begin{eqnarray}
V^{\rm eff}_{\bf k, k', q} \approx g_{\rm sp}^{\rm zb} \,  \sqrt{|{\bf k}||{\bf k'}|}\, \sin(\varphi/2) ,
\label{Vp_opt2}
\end{eqnarray}
where $\varphi$ is the angle between ${\bf k}$ and ${\bf k'}$ and anomalous vertex is the same with 
$\sin\!\rightarrow\!\cos$. The only difference of the effective coupling in (\ref{Vp_opt2}) from the 
coupling  to the optical phonon proposed in (\ref{Vopt}) is the additional angular dependence. It is easy to see
that when calculating the scattering probabilities with $|V|^2$ the latter will be averaged to an additional factor $1/2$ and 
thus corresponds to a simple change of the effective coupling constant.

\subsubsection{Optical phonons}

To introduce optical phonons into the spin-lattice coupling model one needs
to depart from the Bravias picture of (\ref{Hsph0}). For that, keeping in mind cuprates, 
mutual displacement of copper ions 
$\Delta{\bf U}_{i,\delta}\!\equiv\!\left({\bf U}_{i+\delta}\!-\!{\bf U}_i\right)$ 
should now be recognized as involving the complete set of phonon normal modes.
That is, the summation over $\ell$ in (\ref{Uq}) should now be treated as involving 
not only  three acoustic branches of the Bravias lattice, but the entire group 
of optical modes as well. In that sense, the general form of the spin-coupling Hamiltonian 
in (\ref{Hsph1}) and the entire consideration of it remains valid with the magnon-phonon
coupling constants and optical phonon energies used accordingly. 

It is then clear that the small-${\bf q}_\parallel$ scatterings involving
optical branches should be less important than the ones involving acoustic modes, while the 
large-${\bf q}_\parallel$ do not carry additional smallness of $|{\bf q}_\parallel|\!\sim\!T/v$ 
and should be important. Therefore, a consideration of the leading effect of the 
optical-phonon coupling is very similar to that of the zone-boundary phonon above.
Thus, without going into consideration of the details of the crystal structure of specific materials
and neglecting nonessential angular dependence  similarly to the  cases considered above yields 
\begin{eqnarray}
V^{\rm eff}_{\bf k, k', q} \approx g_{\rm sp}^{\rm opt} \,  \sqrt{|{\bf k}||{\bf k'}|}\, ,
\label{Veffopt}
\end{eqnarray}
with $g_{\rm sp}^{\rm opt}\!\approx\!4\lambda S/\sqrt{2m\omega_0}$ where $\omega_0$ is the energy
of the optical branch. Thus, the form of the magnon-phonon interaction proposed in (\ref{Vopt}) is verified.
While, obviously, the coupling strength must depend on the type of the optical model involved,
one can suggest\cite{Ronnow14}  the so-called ``stretching mode'' at high energies as 
a strong candidate for a significant spin-phonon coupling.

\section{Magnitude of spin phonon coupling\label{sub:Magnitude-of-spin}}
\label{appD}

 All three effective 
magnon-phonon coupling constants, to acoustic, to  zone-boundary,
and to optical modes, introduced in Eqs.~(\ref{Vmagph3}), (\ref{Vp_opt1}), and (\ref{Veffopt}), respectively,
have very similar structure: $g_{\rm sp}\!\propto\!\lambda/\sqrt{2m\Theta_D}$ with a coefficient of order of unity. 
Here $\lambda\!=\! a\partial J/\partial r \!\approx\!\gamma J$ is the response of the superexchange constant to the
atomic displacement. It has been argued\cite{Bramwell90} that because the superexchange is very sensitive to
the interatomic distance, the typical values of $\gamma$ are $\sim\! 10-20$. However, this largeness is offset by the
smallness of a characteristic scale associated with phonons,\cite{Bramwell90} 
$1/\sqrt{m\Theta_D}\!\sim\! 1/100$. This is, actually, the same parameter that characterizes the 
smallness of the typical magnitude 
of the zero-point atomic displacement relative to the interatomic distance\cite{Ziman} or 
the smallness of the typical velocity of an atom in a lattice relative to the sound velocity.
Thus, the physical range of the magnon-phonon constants is $g_{\rm sp}/J\!\sim\!0.1$. 

To make a closer estimate of the spin-phonon coupling constants as related to the cuprates, 
we use $m=m_{\rm Cu}$ and a typical phonon 
Debye energy $\Theta_{D}=400$K.  The largest uncertainty is in the value of 
$\partial J/\partial r=\gamma J/a$ with $\gamma\simeq
O(3\ldots14)$ \cite{Ronnow14,Takigawa}. Therefore
\begin{equation}
\frac{4S\lambda\hbar}{\sqrt{2ma^2k_B\Theta_{D}}}\equiv gJ\sim
O(0.05\dots0.3)\,J
\,,\label{21}
\end{equation}
where we have used $a\simeq3.8\mathbf{\textrm{\AA}}$, as in
La$_{2}$CuO$_{4}$.  This defines the dimensionless spin-phonon coupling constant $g$ and
demonstrates that spin-phonon coupling in the cuprates,  while significant,
is still within the bounds to justify the use of perturbation theory.

\section{Verification of the ``effective phonon DoS'' approach}
\label{appB}

Here we complement the discussion of Secs.~\ref{Sec_opt}  and \ref{Sec_ac}
by providing  more verifications of the effective DoS approach  advocated in this work. 

\subsubsection{Optical phonons}

\begin{figure}[tb]
\includegraphics[width=0.999\columnwidth]{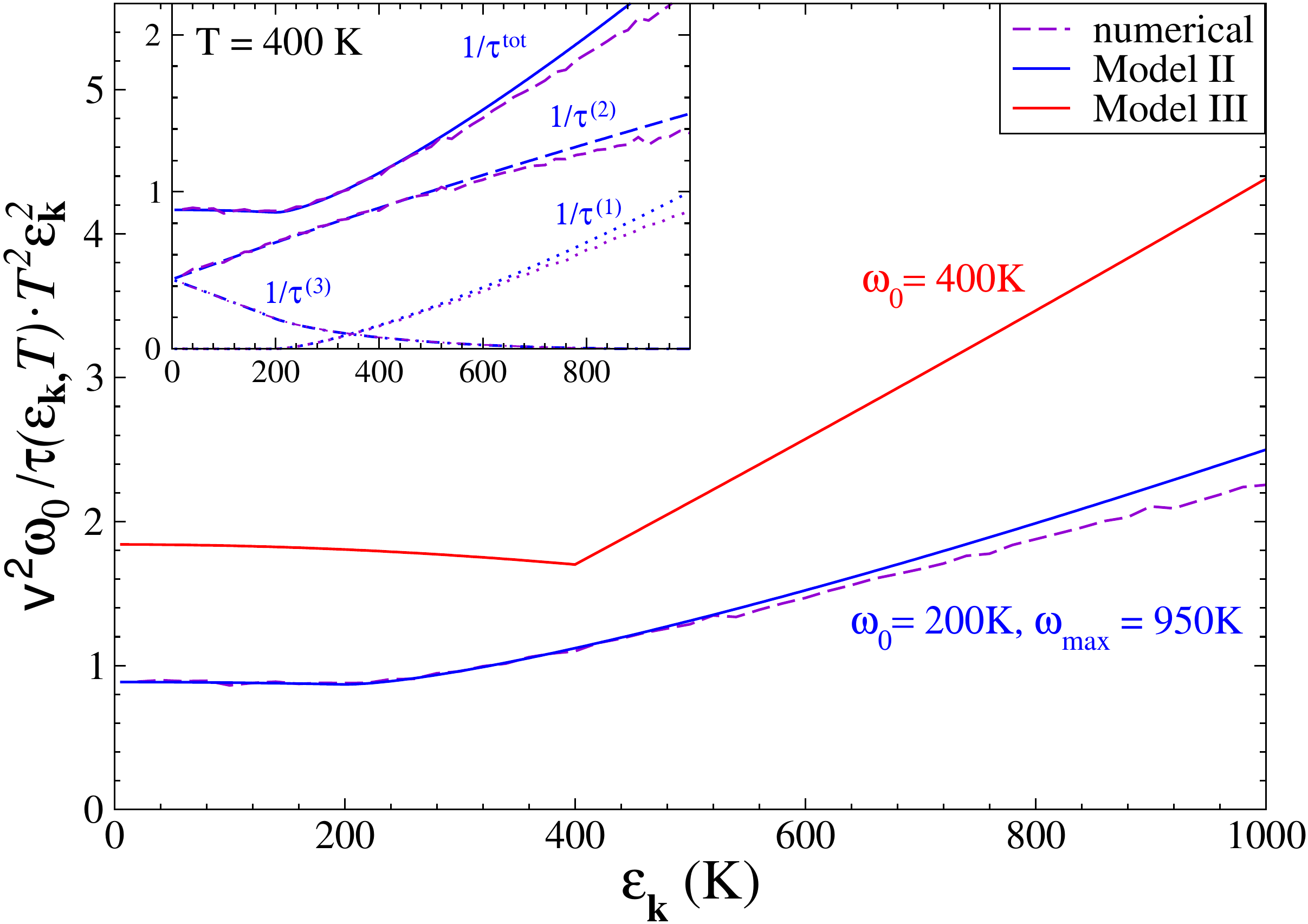}
\caption{(Color online)\ Magnon relaxation rate on the optical phonons, Eq.~(\ref{3tau}), vs $\varepsilon_{\bf k}$
for a representative $T\!=\!400$K and for effective phonon DoS Models II and III in Eqs.~(\ref{phDos2}) and 
(\ref{phDos3}). The results are normalized to the high-temperature asymptotic behavior of the 
phonon-absorption term $\tau_{\bf k}^{(2)}$, Eq.~(\ref{est_tau1}), 
[$T^2\varepsilon_{\bf k}^2/v^2\omega_0$]. 
Parameters are indicated in the graph. 
The vertical axis is in units of $(g_{\rm sp}^{\rm opt}/v)^2$. 
Dashed line is the result of the direct numerical integration in Eq.~(\ref{1tau}) for the dispersive optical phonon
in Fig.~\ref{sketch} without the approximation of Eq.~(\ref{approx}). 
Inset shows individual contributions of the three diagrams in Fig.~\ref{diagrams} 
for both the effective DoS approach (\ref{3tau}) with the Model II and the direct numerical integration.}
\label{Fig1tau_Ek}
\vskip -0.4cm
\end{figure} 

For the ``Model III'' in Eq.~(\ref{phDos3}) (flat phonon mode) of the effective phonon DoS the integral in Eq.~(\ref{3tau}) 
is trivially removed and the relaxation rate is given by a compact analytical expression
\begin{eqnarray}
&&\frac{1}{\tau_{\bf k}}\approx \left(\frac{g_{\rm sp}^{\rm opt}}{v}\right)^2 
\left(\frac{\varepsilon_{\bf k}}{v^2}\right)
\label{tauIII}\\
&&
 \times\Big\{\Theta\big(\varepsilon_{\bf k}-\omega_0\big)
 \left(\varepsilon_{\bf k}-\omega_0\right)^2\Big(n(\omega_0)+n(\varepsilon_{\bf k}-\omega_0)+1\Big)
\nonumber\\
&& \phantom{\frac{1}{\tau_{\bf k}}\Big\{\Theta\big(\varepsilon_{\bf k}-\omega_0\big)}
+\left(\varepsilon_{\bf k}+\omega_0\right)^2\Big(n(\omega_0)-n(\varepsilon_{\bf k}+\omega_0)\Big)
\nonumber\\
&& \phantom{\frac{1}{\tau_{\bf k}}\Big\{}
+\Theta\big(\omega_0-\varepsilon_{\bf k}\big)
\left(\omega_0-\varepsilon_{\bf k}\right)^2\Big(n(\omega_0-\varepsilon_{\bf k})-n(\omega_0)\Big)\Big\}.
\nonumber
\end{eqnarray}
An identical expression can be obtained directly from Eq.~(\ref{1tau}) for the optical 
phonon energy $\omega_{\bf q}\!=\!\omega_0$ and linearized magnon energy and magnon-phonon vertex in 
(\ref{Ek}) and (\ref{Vopt}). Thus, in this case, the effective phonon DoS approach is exact.

Fig.~\ref{Fig1tau} shows the expected activated and high-temperature asymptotic behavior of 
the scattering rate vs $T$ from (\ref{est_tau2}).
Fig.~\ref{Fig1tau_Ek} demonstrates the validity of another aspect of the  asymptotic consideration in 
Eq.~(\ref{est_tau1}): the dependence of $1/\tau_{\bf k}$ on $\varepsilon_{\bf k}$ at $T\!>\!\omega_0$. 
The results in Fig.~\ref{Fig1tau_Ek} are normalized to the high-temperature asymptotic behavior of the 
phonon-absorption term $1/\tau_{\bf k}^{(2)}$ in Eq.~(\ref{est_tau1}), 
$T^2\varepsilon_{\bf k}^2/v^2\omega_0$.
Clearly, for $\varepsilon_{\bf k}\!<\!T$ such a behavior is confirmed (finite intersect of the vertical axis),
while  the phonon-emission term, $1/\tau_{\bf k}^{(1)}$,  carries higher power of $\varepsilon_{\bf k}$, 
also in agreement with Eq.~(\ref{est_tau1}). 
In addition, the results of a direct 3D numerical integration in Eq.~(\ref{1tau}) 
for the dispersive optical phonon in Fig.~\ref{sketch}, 
$\omega_{\bf q}\!\approx\! \omega_0\!+\! \alpha{\bf q}^2$, 
without the approximation of Eq.~(\ref{approx}) are 
shown by the dashed line. Inset shows individual contributions of three diagrams in Fig.~\ref{diagrams} 
for both the effective DoS approach (\ref{3tau}) with the Model II and the direct numerical integration.
One can see a very close agreement of the effective DoS method with the direct numerical integration in (\ref{1tau}), 
which is achieved at a fraction of numerical cost as the former approach requires only 
a 1D integration in Eq.~(\ref{3tau}).

\begin{figure}[tb]
\includegraphics[width=0.999\columnwidth]{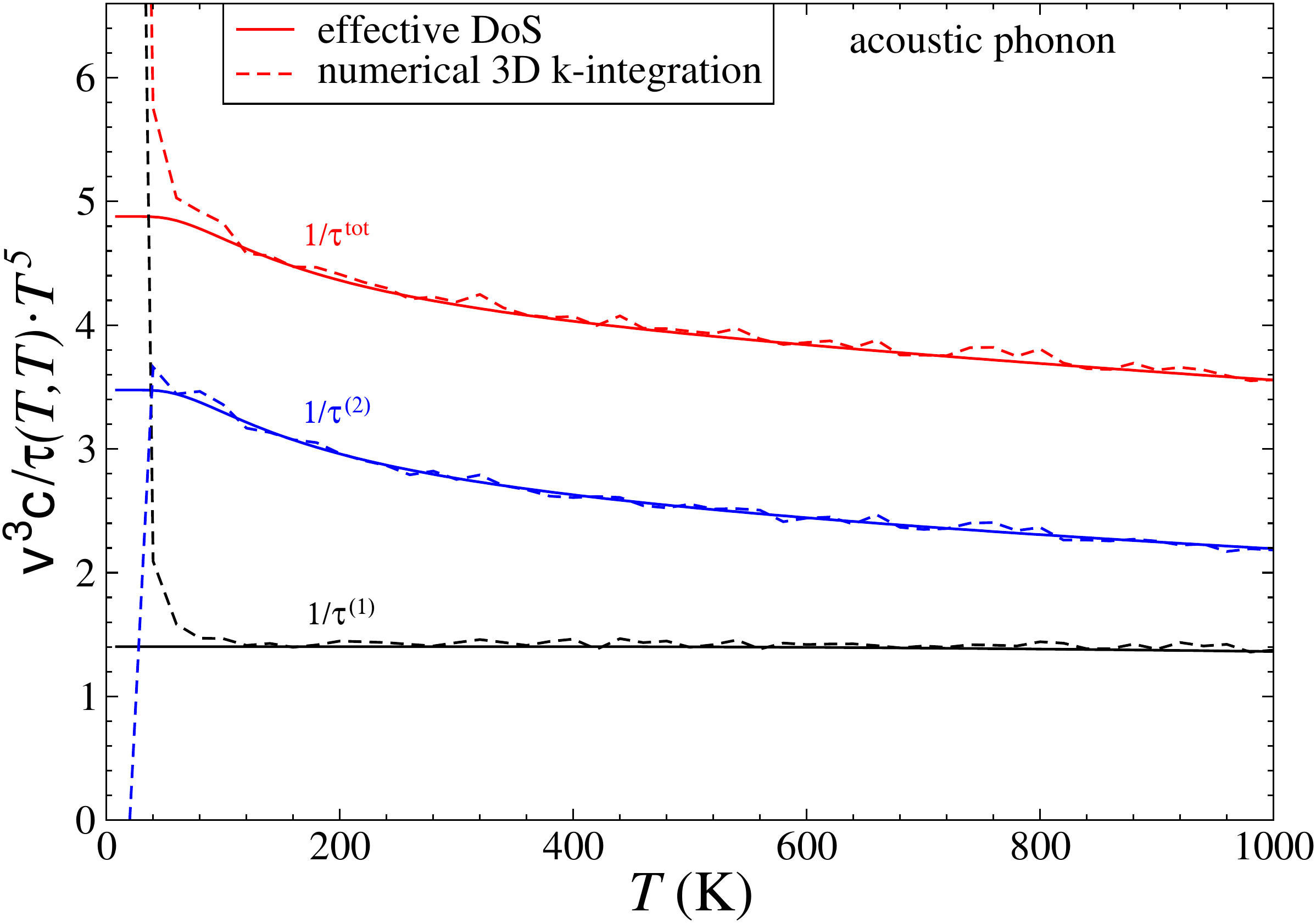}
\caption{(Color online)\ $T$-dependence of the magnon relaxation rate on the acoustic phonons, Eq.~(\ref{4tau}), 
for $\varepsilon_{\bf k}\!=\!T$ and using effective phonon DoS model in Eq.~(\ref{phDos4}).
The results are normalized to the asymptotic behavior, Eq.~(\ref{est_tau4}), 
[$T^5/v^3\omega_0$]. Parameters are as discussed in text, Debye energy $\Theta_D\!=\!400$K. 
The vertical axis is in units of $(g_{\rm sp}^{\rm ac}/v)^2$. Individual contributions of the first two 
terms in (\ref{4tau})  [diagrams in Fig.~\ref{diagrams}(a) and (b)] are indicated. 
Dashed lines are the result of the direct numerical 3D integration in Eq.~(\ref{1tau}) for the acoustic phonon
without the approximation of Eq.~(\ref{approx}). }
\label{Fig2tau}
\vskip -0.4cm
\end{figure} 
\begin{figure}[tb]
\includegraphics[width=0.999\columnwidth]{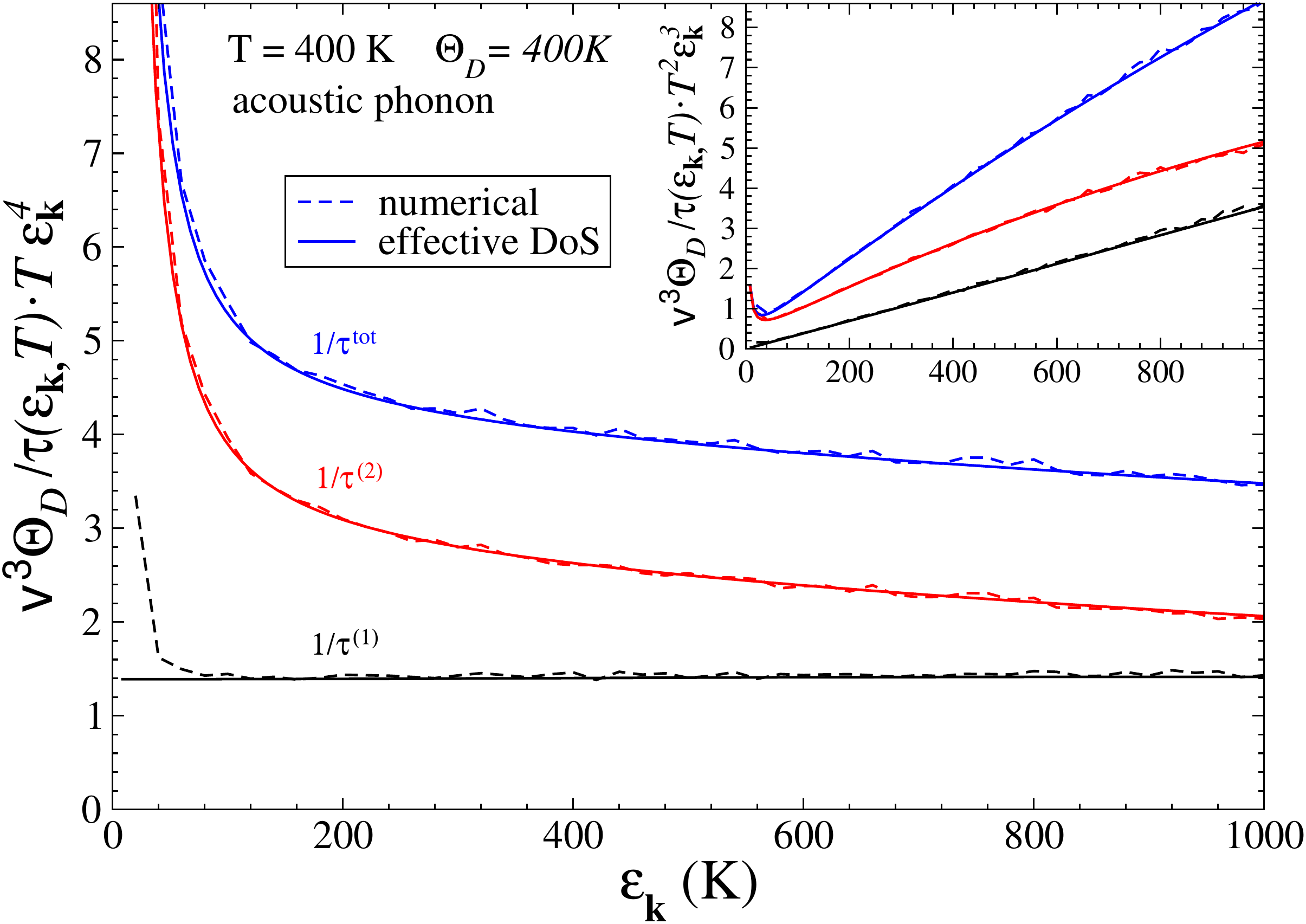}
\caption{(Color online)\ 
Magnon relaxation rate, Eq.~(\ref{4tau}), vs $\varepsilon_{\bf k}$
for a representative $T\!=\!400$K and for effective phonon DoS model  in Eq.~(\ref{phDos4}). 
The results are normalized to the  asymptotic behavior, Eq.~(\ref{est_tau3}), 
[$T\varepsilon_{\bf k}^4/v^3\Theta_D$]. 
Inset: Same with the normalization by $T^2\varepsilon_{\bf k}^3/v^3\Theta_D$.}
\label{Fig2tau_Ek}
\vskip -0.4cm
\end{figure} 

\subsubsection{Acoustic phonons}

With the help of our Figs.~\ref{Fig2tau} and \ref{Fig2tau_Ek} 
we provide a demonstration of the accuracy and numerical efficiency 
of the effective phonon DoS approach.
In them the $T$- and 
$\varepsilon_{\bf k}$-dependencies of the relaxation rate by acoustic phonons from (\ref{4tau})
are compared with an explicit numerical 3D integration in (\ref{1tau}) with the magnon-phonon vertex
from (\ref{Vac}) and without the use of approximation (\ref{approx}). These figures also offer an 
additional confirmation of the asymptotic trends of Eqs.~(\ref{est_tau4}) and (\ref{est_tau3}).

In Fig.~\ref{Fig2tau}, the $T$-dependence is shown for the relaxation rate Eq.~(\ref{4tau}) for
$\varepsilon_{\bf k}\!=\!T$, i.e., on ``thermal shell''. The results are normalized to $T^5/v^3\Theta_D$ 
to make the  asymptotic behavior of (\ref{est_tau4}) apparent. The vertical axis is in 
units of $(g_{\rm sp}^{\rm ac}/v)^2$. We also show individual contributions of the first two
terms in (\ref{4tau}), while the third  is negligible as discussed above.  
The results of a direct 3D numerical integration in Eq.~(\ref{1tau}) 
for acoustic phonon  (within the Debye approximation) without the approximation of Eq.~(\ref{approx}) are 
shown by  dashed lines. One can see a very close quantitative agreement of the effective DoS 
method with the direct numerical integration. At very low $T$, the direct numerical procedure 
becomes unreliable due to  very small ${\bf k'}$- and ${\bf q}$-space of integration relevant for the scattering.
Again, the high-accuracy results of the effective DoS approach are achieved at a fraction of the 
numerical cost of the direct integration.

Fig.~\ref{Fig2tau_Ek} shows the dependence of $1/\tau_{\bf k}$ on $\varepsilon_{\bf k}$ at a representative 
$T\!=\!400$K. The results in Fig.~\ref{Fig2tau_Ek} are normalized to the asymptotic behavior in 
Eq.~(\ref{est_tau3}), $T\varepsilon_{\bf k}^4/v^3\Theta_D$.
Clearly, for $1/\tau^{(1)}$  such an asymptote is essentially precise for all the energies, while 
for the phonon-absorption term substantial deviations occur at lower energies, the feature also emphasized
in the inset of Fig.~\ref{Fig2tau_Ek}. This can be 
analyzed by a more careful asymptotic treatment of Eq.~(\ref{4tau}) in the 
$\varepsilon_{\bf k}\!\ll\!T$ regime, which show that the smallness of the subleading terms by $c/v$ gets
compensated by the largeness of $T/\varepsilon_{\bf k}$ for small enough $\varepsilon_{\bf k}$, and the 
ultimate asymptotic behavior of this term in the 
$\varepsilon_{\bf k}\!\ll\!T$ regime is $1/\tau^{(2)}\!\sim\!T^3\varepsilon_{\bf k}^2/v^4$.
As in Fig.~\ref{Fig2tau},  Fig.~\ref{Fig2tau_Ek} shows a very close quantitative agreement of the effective DoS 
method with the direct numerical integration.




\begin{thebibliography}{99}
\bibitem{Theta}  In practice, the ``pseudo''-step-function $\widetilde{\Theta}(x)$  can be chosen as 
$\widetilde{\Theta}(x_i)\!=\!1\!-\!1/(e^{x_i}+1)$, where $x_i\!=\!(T-\omega_{0,i}/2)/\Delta_i$ and 
$\Delta_i$ defines the ``width'' of the 
exponential turn-on for the scattering on the $i$th optical mode.
\end{thebibliography}
\end{document}